%% file: acl_latex.tex
\pdfoutput=1
\documentclass[12pt]{article}

\usepackage[T1]{fontenc}
\usepackage[utf8]{inputenc}
\usepackage{times}
\usepackage{latexsym}
\usepackage{microtype}
\usepackage{float}
\usepackage{inconsolata}
\usepackage{graphicx}
\usepackage{amssymb}
\usepackage{amsmath}
\usepackage{booktabs}
\usepackage{subcaption}
\usepackage{multirow}
\usepackage{hyperref}
\usepackage{xcolor}
\usepackage[mode=text, group-separator=\text{\ }]{siunitx}
\usepackage{todonotes}
\usepackage{listings}
\usepackage{tcolorbox}
\tcbuselibrary{listings, breakable}

\usepackage{natbib}
\usepackage[margin=0.8in]{geometry}
\newtcolorbox{PromptBox}[1][]{%
	breakable,
	colback=gray!5,
	colframe=black!40,
	title=#1,
	fonttitle=\bfseries,
	boxrule=0.4pt,
	arc=2pt,
	top=2pt,
	bottom=2pt,
	left=4pt,
	right=4pt,
	width=\textwidth,
	listing only,
	listing options={
		basicstyle=\ttfamily\small,
		breaklines=true,
		breakatwhitespace=false,
		showstringspaces=false,
		columns=fullflexible,
		keepspaces=true,
	},
}

\expandafter\def\expandafter\normalsize\expandafter{%
	\normalsize%
	\setlength\abovedisplayskip{4pt}%
	\setlength\belowdisplayskip{4pt}%
	\setlength\abovedisplayshortskip{0pt}%
	\setlength\belowdisplayshortskip{2pt}%
}

\title{\bfseries AutoBool: An Reinforcement-Learning trained LLM for Effective Automated Boolean Query Generation for Systematic Reviews}

\author{
	Shuai Wang$^{1}$ \quad
	Harrisen Scells$^{2}$ \quad
	Bevan Koopman$^{1,3}$ \quad
	Guido Zuccon$^{1}$ \\
	\normalsize $^1$The University of Queensland, Brisbane, Australia \\
	\normalsize $^2$University of Tübingen, Tübingen, Germany \\
	\normalsize $^3$CSIRO, Brisbane, Australia \\
	\normalsize \texttt{shuai.wang2@uq.edu.au, harrisen.scells@uni-tuebingen.de} \\
	\normalsize \texttt{bevan.koopman@csiro.au, g.zuccon@uq.edu.au}
}

\date{}

\begin{document}
	\maketitle
	
	\begin{abstract}
	We present \textbf{AutoBool}, a reinforcement learning (RL) framework that trains large language models (LLMs) to generate effective Boolean queries for medical systematic reviews. Boolean queries are the primary mechanism for literature retrieval in this domain and must achieve high recall while maintaining reasonable precision---a challenging balance that existing prompt-based LLM approaches often struggle to achieve.
	A major limitation in this space is the lack of high-quality ground-truth Boolean queries for each topic, which makes supervised fine-tuning impractical. \textbf{AutoBool} addresses this challenge by using RL to directly optimize query generation with retrieval  measures, without requiring target queries.
	To support this effort, we create and release the largest dataset of its kind: \num{65 588} topics in total for training and evaluating the task of automatic Boolean query formulation.
	
	Experiments on our new dataset and two established datasets (CLEF TAR and Seed Collection) show that AutoBool significantly outperforms zero-shot/few-shot prompting and matches or exceeds the effectiveness of much larger GPT-based models (e.g., GPT-4o, O3) using smaller backbones. It also approaches effectiveness of expert-authored queries while retrieving 10–16 times fewer documents. Ablation studies reveal the critical roles of model backbone, size, decoding temperature, and prompt design. Code and data are available at~~\url{https://github.com/ielab/AutoBool}.
	\end{abstract}
	
	\input{sections/introduction.tex}
	\input{sections/related_works.tex}
	\input{sections/dataset.tex}
	\input{sections/method.tex}
	\input{sections/experiment_setup.tex}
	\input{sections/result.tex}

	\input{sections/ablations.tex}
	\input{sections/conclusion.tex}

	\appendix
	\input{sections/appendix.tex}

	\bibliography{custom}
	
\end{document}

%% file: sections/introduction.tex
\section{Introduction}
Systematic reviews are essential tools in evidence-based medicine, providing comprehensive and unbiased summaries of scientific knowledge across medicine and social sciences. At the heart of these reviews lies a deceptively complex task: the formulation of Boolean search queries capable of retrieving all relevant literature (high recall) without overburdening researchers with non-relevant results (high precision). A well-formulated Boolean query directly affects the cost, efficiency, and reproducibility of the review process.

Large language models (LLMs) have been explored as tools for automatically generating Boolean queries from an initial research question or topic~\cite{wang2023can, wang2025reassessing,staudinger2024reproducibility}. While conceptually easy to use, prompt-based LLM methods have shown major limitations---often retrieving far fewer relevant studies than expert-crafted queries (low recall), well below the thresholds typically required for systematic reviews (e.g., 10--40\% recall instead of 80--90\%)~\cite{wang2025reassessing}. These limitations highlight the need for new methods that go beyond zero-shot prompting and instead are optimized to generate queries based on retrieval effectiveness.

A natural alternative to prompting is supervised fine-tuning on example queries. However, this is impractical for Boolean query generation, where no single high-quality ground-truth query exists. Expert-crafted queries are often inconsistent, and sub-optimal~\cite{scells2018generating}. Further, existing datasets are too small: fewer than 200 training pairs, many sourced from Cochrane~\cite{kanoulas2018clef, wang2022seed}.%
\footnote{Cochrane does not support direct API access; prior work translates these queries into MEDLINE format for PubMed execution~\cite{wang2025reassessing}.}

We propose a reinforcement learning (RL) framework called \textbf{AutoBool} to train LLMs for Boolean query generation, enabling direct optimization for retrieval effectiveness. Rather than relying on handcrafted prompts or static templates, our model learns to balance recall and precision through feedback from document retrieval. To support this task, we construct a large-scale training dataset of \num{32794} systematic reviews mined from the PubMed Central (PMC) Open Access corpus, and introduce a retrieval-grounded reward function aligned with the screening goals of systematic review creation.
We further release a large-scale evaluation dataset comprising \num{32794} topics with high-quality relevance labels---an order of magnitude larger than prior benchmarks such as CLEF TAR and the Seed Collection,%
\footnote{Which contain 118 and 40 topics overall respectively.} 
and with a smaller PubTemp set designed to be free from data leakage. Figure~\ref{fig:overview} illustrates our complete pipeline, from dataset construction to RL training.

Our experiments show that \textbf{AutoBool}-trained models substantially outperform prompt-based zero-shot baselines, narrow the gap with expert-crafted queries, and even exceed the effectiveness of much larger commercial LLMs such as GPT-4o and O3 in high-recall retrieval scenarios, despite relying on significantly smaller backbones.

%% file: sections/related_works.tex
\begin{figure*}[htbp]
	\centering
	\includegraphics[width=0.95\textwidth]{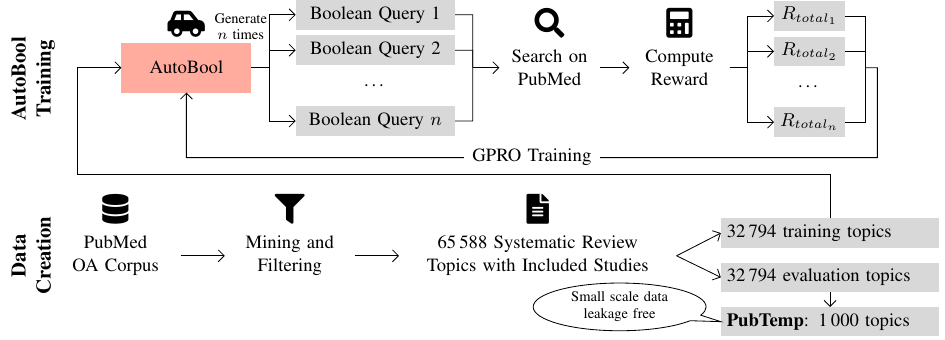}
	\caption{Overview architecture of dataset creation and AutoBool training.}
	\label{fig:overview}
\end{figure*}

\section{Related Work}
\label{sec:related_work}
\textbf{Traditional Approaches to Boolean Query Formulation.}
Previous work on automated Boolean query formulation followed two main paradigms: the \emph{objective method} and the \emph{conceptual method}~\cite{scells2020conceptual, scells2020objective}. The objective method aims to replicate known included studies by expanding on initial seed sets using keyword co-occurrence, term frequency analysis, or relevance feedback~\cite{hausner2012routine}. These approaches prioritize recall but sacrifice interpretability and precision. The conceptual method involves manually identifying key aspects (population, intervention, outcome) and converting them into Boolean expressions. While more interpretable and aligned with expert workflows, they require substantial manual effort and domain expertise~\cite{clark2013systematic}.

\textbf{LLM-Based Query Generation.}
LLMs have recently been explored for automating Boolean query generation from topic descriptions. While they can produce syntactically valid queries, they often require multiple attempts, and their recall remains well below expert-authored queries~\cite{wang2025reassessing}. These limitations highlight the need for trainable methods beyond static prompting. Supervised fine-tuning is not well-suited here, as there is no single high-quality ground-truth query per topic: expert-authored queries are manually refined, inconsistent, and often suboptimal.

\textbf{Reinforcement Learning for LLM Optimization.}
RL offers a flexible way to optimize models based on task-level feedback without requiring gold-standard targets. It has emerged as a promising approach for optimizing LLMs on non-differentiable objectives such as factual accuracy, preference alignment, and retrieval quality~\cite{zhang2020deep, zhuang2025rank, nguyen2024reinforcement}. Notably, Group Relative Policy Optimization (\textbf{GRPO}) has shown strong performance in LLM fine-tuning scenarios, including DeepSeek-R1~\cite{deepseekai2025deepseekv3technicalreport, guo2025deepseek}.

%% file: sections/dataset.tex
\section{Dataset Creation}
\label{sec:dataset}
To support both training and large-scale evaluation of systematic review Boolean query generation, we construct a new dataset based on full-text systematic reviews from the PubMed Central (PMC) Open Access (OA) subset~\cite{pmc_open_access_subset}. This subset is license-compatible with commercial use and substantially larger than existing public benchmarks.%
\footnote{Dataset will be made available on~\href{https://www.example.com}{Huggingface}.}

\subsection{Data Source and Extraction}
We begin by extracting all articles labeled with the publication type \texttt{systematic review} from PubMed Central (PMC) Open Access (OA) subset, resulting in a total of \num{75676} systematic review topics. For each topic, we parse the full PMC XML and extract the PMIDs cited in the results section. These cited references are treated as the included studies (i.e., the gold-standard relevant set) for that review. After filtering for availability of cited PMIDs, we retain \num{65600} usable topics.

\subsection{Benchmark Integrity}
To prevent data leakage from prior evaluation sets, we manually remove any topics overlapping with the CLEF TAR or Seed Collection~\cite{kanoulas2018clef, wang2022little}. This resulted in the exclusion of 12 topics, yielding a final dataset of \num{65588} unique systematic reviews topics.

\subsection{Temporal Train-Test Split}
To prevent temporal leakage and better simulate real-world deployment, we split the dataset chronologically based on publication date:
\begin{itemize}\setlength\itemsep{0ex}
	\item \textbf{Training set:} \num{32794} topics published between \texttt{2000-07-06} and \texttt{2021-10-30}.
	\item \textbf{Test set:} \num{32794} topics published between \texttt{2021-10-31} and \texttt{2025-03-01}.
	\item \textbf{PubTemp} (\textbf{PubMed Temporal}) \textbf{Set:} 1,000 randomly sampled topics published after \texttt{2024-11-01}.
\end{itemize}

This temporal split reflects real-world usage: an information specialist formulates a Boolean query to retrieve studies published up to that point. The model, trained only on earlier topics, must generalize to future, unseen ones. Chronological partitioning also supports continual evaluation, enabling future research on adapting to evolving terminology, intervention types, and publication trends.

We introduce the \textbf{PubTemp} set to enable fair, out-of-distribution evaluation by ensuring topics were unseen during LLM pretraining. Since Qwen3 (the primary model used) has an October 2024 knowledge cutoff,%
\footnote{While there is no officially published knowledge cutoff for Qwen3, we confirmed via direct model queries that its knowledge extends up to October 2024.} 
we select topics published after November 1, 2024, to minimize overlap. To keep evaluation feasible, we randomly sample \num{1000} such topics to avoid the API time cost of issuing too many PubMed API queries and reducing the expense of evaluating commercial models (e.g., generating queries for \num{1000} topics using O3 costs around~\$50). The PubTemp split will be released with our dataset to support reproducibility.

%% file: sections/method.tex
\section{Method}
We train an LLM to generate Boolean queries using GRPO to optimize retrieval effectiveness as detailed in Section~\ref{sec:related_work}. The central challenge in applying GRPO lies in designing an appropriate reward function. For systematic reviews, the effectiveness of a Boolean query can be evaluated by executing the query on a document collection (e.g., PubMed) and comparing the retrieved documents against a gold-standard set of included studies. This enables computation of retrieval metrics such as recall and precision which form the basis of the reward.

\subsection{Reward Design}
\label{sec:retrieval_reward}
The total reward consists of three components: formatting correctness, syntactic validity, and retrieval effectiveness.

\paragraph{Formatting Reward.}
The formatting reward \( R_{\text{format}} \) checks whether the output follows expected structural conventions (e.g., quoted terms, capitalized Boolean operators):
\begin{equation}\label{eq:format_reward}
	R_{\text{format}} =
	\begin{cases}
		10 & \text{if format is correct} \\
		-10 & \text{otherwise}
	\end{cases}
\end{equation}
\paragraph{Validity Reward.}
The validity reward \( R_{\text{validity}} \) ensures that the generated Boolean query can be both syntactically parsed and successfully executed in a retrieval system:
\begin{equation}\label{eq:validity_reward}
	R_{\text{validity}} =
	\begin{cases}
		10 & \text{if query is valid} \\
		-10 & \text{if query is invalid}
	\end{cases}
\end{equation}
A query is considered valid if it passes two checks:
(1) it must be syntactically correct according to a custom Boolean query parser that verifies structural elements such as balanced parentheses and proper use of logical operators; and
(2) it must return at least one result when executed via PubMed, and returns fewer than a maximum threshold of 200,000 documents.%
\footnote{This threshold is enforced to ensure query efficiency and system responsiveness.}
Queries that fail either check are treated as invalid and receive a penalty.

\paragraph{Retrieval Reward.}
The retrieval reward is designed to support recall-oriented Boolean query generation, reflecting the priorities of systematic reviews: retrieve as many relevant studies as possible while minimizing screening burden. It balances a direct reward for recall and a recall-modulated reward for precision.

The reward function satisfies three core properties: 
(1) \textbf{recall must be prioritized}, as high recall is essential in systematic reviews~\cite{straube2021recall};
(2) \textbf{precision should matter more as recall increases}, since reducing irrelevant results becomes valuable once coverage improves; and 
(3) \textbf{early precision gains should be emphasized}, as small improvements from very low precision levels bring significant practical benefit.

To implement this, we apply two mechanisms: 
(1) weighting precision by \( r^\alpha \) to reduce its impact at low recall; and 
(2) applying a logarithmic transformation to increase sensitivity when precision is low. The resulting reward is:
\begin{equation}\label{eq:reward_function}
	\resizebox{0.5\columnwidth}{!}{$\displaystyle
		F(r, p) = \underbrace{M \cdot r}_{\text{recall reward}} + \underbrace{M \cdot r^\alpha \cdot \log_{1 + s}(1 + s \cdot p)}_{\text{precision reward}}
		$}
\end{equation}
where \( r \) is recall, \( p \) is precision, \( s = 100 \) is a smoothing constant, \( \alpha \geq 0 \) controls precision weighting, and \( M \) is a global scaling factor.
The complete retrieval reward is defined as:
\begin{equation}\label{eq:retrieval_reward}
	\resizebox{0.5\columnwidth}{!}{$\displaystyle
		R_{\text{retrieval}}(r, p, |D|) =
		\begin{cases}
			-20 &\!\text{if } |D| = 0 \\
			-5 &\!\text{if } r = 0 \land p = 0 \\
			F(r, p) &\!\text{otherwise.}
		\end{cases}
		$}
\end{equation}
The choice of \( \alpha \) significantly influences retrieval behavior. When \( \alpha = 0.5 \), the reward is \textit{weakly recall-oriented}, allowing precision to contribute earlier in the learning process. At \( \alpha = 1 \), the behavior becomes \textit{moderately recall-oriented}, offering a balanced trade-off between recall and precision. Setting \( \alpha = 2 \) results in a \textit{strongly recall-oriented} reward, where precision only meaningfully contributes once high recall has been achieved.
\paragraph{Total Reward.}
The final reward combines all components: 
\begin{equation}\label{eq:total_reward}
	R_{\text{total}} = R_{\text{format}} + R_{\text{validity}} + R_{\text{retrieval}}
\end{equation}

\subsection{Training Procedure}
We fine-tune a pretrained LLM using policy optimization to maximize \( R_{\text{total}} \). At each training step, the model is prompted with a systematic review topic and generates a Boolean query. This query is executed on a document collection, and its reward is computed. Gradients from GRPO are used to update the model, encouraging generation of valid, well-structured, and high-recall queries over time.

\subsection{Prompting Strategies}
We investigate four prompting strategies, each designed to elicit different forms of reasoning. No Reasoning \textbf{(N.R)} and Free-text Reasoning \textbf{(R)} provide essential instructions about what a Boolean query is, its components, requirements and how to use different search fields. Conceptual Reasoning \textbf{(R-con)} and Objective Reasoning \textbf{(R-obj)} are inspired by established approaches in query formulation. These prompts offer more structured, step-by-step guidance on how to decompose the topic and construct the query~\cite{scells2020objective, scells2020conceptual}.

Table~\ref{tab:prompt-types} summarizes these strategies. Full prompt templates are provided in Appendix~\ref{appendix:prompts}. 

\begin{table}[t]
	\centering
	\footnotesize
	\caption{Prompting strategies for Boolean query generation. Full templates are in Appendix~\ref{appendix:prompts}.}
	\label{tab:prompt-types}

	\begin{tabular}{p{0.94\columnwidth}} % use 0.95 instead of full width
		\toprule
		\textbf{Prompt Type and Description} \\
		\midrule
		\textbf{No Reasoning (N.R):} Direct query generation with minimal explanation or structure. \\
		\midrule
		\textbf{Free-text Reasoning (R):} Includes a natural language explanation before query generation. \\
		\midrule
		\textbf{Conceptual Reasoning (R-con):} Uses structured decomposition (Population, Intervention, Outcome) to scaffold the query. \\
		\midrule
		\textbf{Objective Reasoning (R-obj):} Simulates a relevant abstract and extracts key terms empirically. \\
		\bottomrule
	\end{tabular}
\end{table}

%% file: sections/experiment_setup.tex
\section{Experimental Setup}

\subsection{Retrieval and Evaluation Protocol}
For all datasets, we use the PubMed Entrez API to execute the generated Boolean queries and retrieve candidate documents by matching on PMIDs~\cite{sayers2010general}. This simulates a realistic literature search workflow, allows us to evaluate query effectiveness in a practical retrieval setting, and follows standard methodology from previous work~\cite{wang2023can, wang2025reassessing}.
We apply the same query validity check protocol described in Section~\ref{sec:retrieval_reward} to detect and reject malformed queries. Queries that fail this check are considered invalid, and the model is prompted to regenerate until a valid query is produced, with a maximum of 10 attempts.%
\footnote{We cap regenerations at 10 to avoid excessive API usage and model inference.}
This validation process aligns with established evaluation practices~\cite{wang2025reassessing}.

We evaluate with measures used in earlier research~\cite{wang2025reassessing}. The \textbf{primary evaluation metrics} are recall, F\textsubscript{3}, and the percentage of queries obtaining recall above 80\% and 90\%. These reflect the high-recall requirements of systematic reviews, where omissions of relevant studies can critically undermine review quality.
As \textbf{secondary metrics}, we report precision, the average number of documents retrieved (to measure screening effort), the average number of regenerations per query, and the success rate under 10 attempts---to capture robustness and generation stability beyond recall-focused evaluation.

\subsection{Model Variants}
Our primary experiments are conducted using models from the Qwen3 family~\cite{yang2025qwen3}. Unless otherwise specified, all trained AutoBool Models are based on \texttt{Qwen3-4B}, which serves as the default backbone throughout our main results. These models are fine-tuned using our reward-driven training framework (Section~\ref{sec:retrieval_reward}) to optimize for systematic review retrieval effectiveness. GRPO is applied to guide query generation behavior based on retrieval performance signals. 
At inference time, we use the same prompt as during training and fix the decoding temperature to 0.6, following the Qwen3 recommendations~\cite{yang2025qwen3}.

\subsection{Evaluation Datasets}
We evaluate our models on three datasets: the \textbf{PubTemp set}, and two established benchmarks: the \textbf{CLEF TAR} collection~\cite{kanoulas2018clef} and the \textbf{Seed Collection}~\cite{wang2022little}.

We follow prior work~\cite{wang2025reassessing} using the CLEF TAR 2017 and 2018 subsets (72 topics total) and the Seed Collection (40 topics). Each topic is defined by experts and paired with a manually curated set of relevant studies. These benchmarks have been widely adopted for evaluating automatic Boolean query generation in the context of systematic review automation~\cite{macfarlane2022search, kusa2023csmed, wang2022neural, stevenson2023stopping, lee2018seed}.

\subsection{Training Parameters}

All models are trained using the GRPO RL algorithm with a retrieval-based reward. We adopt LoRA-based parameter-efficient fine-tuning and use the vLLM backend in colocated mode.
For each prompt the model generates four completions to compute reward scores. We use an effective batch size of 16 via gradient accumulation. Prompt and completion length limits are 768/1024 tokens for non-reasoning prompts and 1024/3072 for reasoning-based prompts.
Unless otherwise noted, we set the training temperature to 1.2; its effect is analyzed in Section~\ref{sec:ablation}. Full training hyperparameters are listed in Table~\ref{appendix:tab-training-details} in Appendix~\ref{appendix:training-details}.

%% file: sections/result.tex
\section{Results}

\input{tables/result-base.tex}

Table~\ref{tab:eval-results-pubmed} summarizes the evaluation results on the PubTemp set. We observe that reinforcement learning substantially improves Boolean query generation performance, especially in recall-critical settings like systematic reviews. 

\subsection{Effectiveness of Reinforcement Learning}
Compared to zero-shot prompting baselines using the same base model (Qwen3-4B), all RL-trained AutoBool models achieve substantially higher effectiveness. Top models obtain 0.70 average recall, with 35\% of queries exceeding 90\% recall, versus 0.35 recall and 6.5\% for zero-shot baselines. We also compare against in-context learning (ICL) with 1-shot, 3-shot, and 5-shot configurations (Appendix~\ref{appendix:icl}): AutoBool achieves 7× improvement over the best ICL result, demonstrating that RL provides optimization signals that cannot be communicated through examples alone.

On secondary metrics, AutoBool models improve precision under the \texttt{N.R} and \texttt{R-obj} prompts, but slightly reduce precision under \texttt{R} and \texttt{R-con}. All trained models retrieve more documents than their zero-shot counterparts: an expected trade-off when optimizing for recall. However, this increase in retrieved set size remains reasonable (well under 1000 on average), and does not significantly increase screening burden compared to the gains in comprehensiveness.

Training also improves generation stability. AutoBool models exhibit consistently higher success rates (near 100\%) and require fewer regenerations than zero-shot baselines, indicating more reliable formatting and syntactic validity—driven by our structured reward components.

\paragraph{Comparison with Larger GPT-Based Models.}
Compared to significantly larger GPT-based models (GPT-4o and O3),%
~\footnote{While parameter counts are proprietary, these models are widely estimated to be ~100× larger than our 4B backbone.}
AutoBool, despite using a much smaller model, achieves higher recall and a higher percentage of queries exceeding high-recall thresholds (e.g., 80\% and 90\%). Regeneration and success rates are also comparable across models. While AutoBool has lower precision and F\textsubscript{3}, this reflects domain-specific optimization: the screening burden difference is modest (O3: 499--603 documents vs AutoBool: 586--773), representing only 5–6 additional hours in a 30+ hour process, while missing studies can invalidate entire reviews. The average number of retrieved documents remains within a practical range, and AutoBool enables local deployment for organizations with API privacy restrictions. These results highlight the effectiveness of retrieval-driven RL training in generating high-recall Boolean queries, even under constrained model capacity.

\paragraph{Effect of Prompt Type.}
Under zero-shot settings, the \texttt{R-con} prompt generally yields the highest recall and F\textsubscript{3} scores (except for F\textsubscript{3} in GPT-4o and recall in O3), suggesting that structured reasoning aids query formulation when no retrieval feedback is available. However, this advantage diminishes after training. For trained AutoBool models, the \texttt{No Reasoning} prompt consistently achieves the best recall and high-recall threshold performance across all settings, while reasoning-based prompts tend to yield higher precision.

We hypothesize two complementary explanations for these trends. First, RL enables the model to internalize task-specific structure and discover its own optimized generation strategy, which may diverge from human-designed decomposition frameworks. Second, Boolean queries are inherently interpretable and self-contained: their logic is fully encoded through syntax and operators. As such, requiring the model to verbalize intermediate reasoning may introduce unnecessary constraints or verbosity once it has learned to generate effective queries end-to-end.

Nonetheless, reasoning-based prompts consistently yield higher success rates and require fewer regenerations than \texttt{No Reasoning}, even after training. This suggests that intermediate reasoning may still offer robustness benefits in more difficult systematic review topics, where generating a valid and well-formed Boolean query is especially challenging. In these cases, explicit reasoning may help the model maintain syntactic and semantic integrity under ambiguity or complexity.

\paragraph{Which \texorpdfstring{$\alpha$}{α} Value Should Be Used?}
We analyze the effect of the \(\alpha\) parameter in the retrieval reward, which adjusts the emphasis on recall over precision. In structured reasoning-based prompts (\texttt{R-con}, \texttt{R-obj}), increasing \(\alpha\) consistently improves recall, as intended. In contrast, results are more mixed for less-structured prompts: \(\alpha = 1\) yields the highest recall with \texttt{N.R}, while \(\alpha = 0.5\) performs best with \texttt{R}. For F\textsubscript{3}, \(\alpha = 1\) generally achieves the best performance across prompts, except in \texttt{N.R}, where \(\alpha = 0.5\) outperforms. Overall, \(\alpha = 1\) offers the most balanced trade-off between recall and precision, making it a strong default. Its effects are also more stable in structured prompts, where recall improves more predictably with higher \(\alpha\).

\begin{figure*}[htbp]
	\centering
	\includegraphics[width=\textwidth]{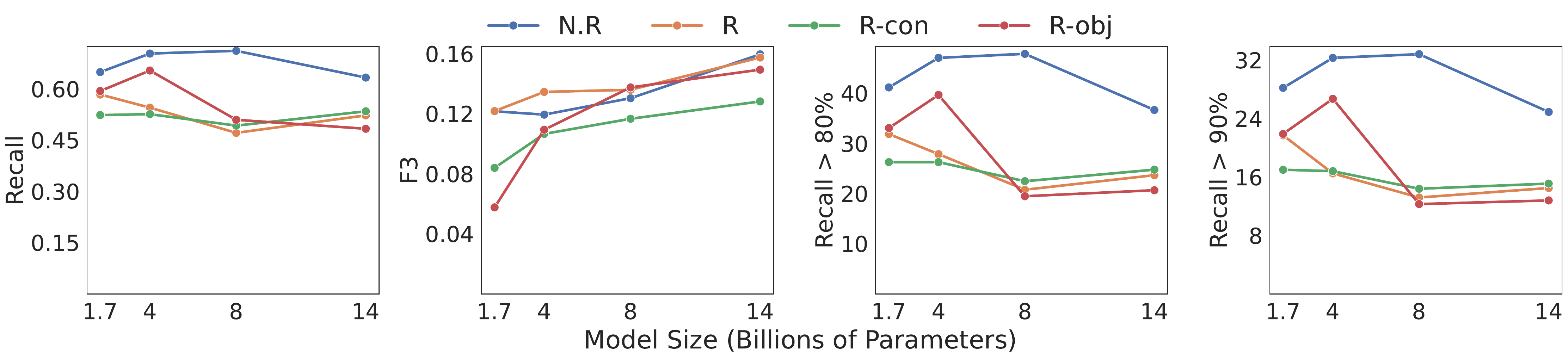}
	\caption{Effect of model size on the effectiveness of Boolean query generation across prompt types on the PubMed Temporal-Cutoff set, all result based on Qwen3 based Models.}
		\label{fig:model-size-comparison}
\end{figure*}

\subsubsection{Generalization on Existing Datasets}

To assess generalization, we evaluate our trained models on two established systematic review benchmarks: CLEF TAR and the Seed Collection (Table~\ref{tab:eval-results-clef} and Table~\ref{tab:eval-results-seed} in Appendix). Both datasets are relatively small and publicly available, raising potential concerns around bias and data leakage.

\textbf{For CLEF TAR,} AutoBool generalizes well, obtaining substantially higher recall and recall-threshold metrics than its zero-shot counterparts. It is also more effective than larger models like GPT-4o and O3 in recall. With \(\alpha = 1\) and the \texttt{No Reasoning} prompt, AutoBool almost matches the recall of expert-crafted queries (within 1\%) while retrieving 17 times fewer documents; yielding improved F\textsubscript{3}, higher precision, and reduced screening effort. It is also more effective than the O1 model from~\citet{wang2025reassessing} in both recall and F\textsubscript{3}, highlighting the benefits of RL optimization.
\textbf{For Seed Collection,} a similar pattern holds. While AutoBool is less effective than expert-written queries, it is significantly more effective than zero-shot baselines and the O1 model from~\citet{wang2025reassessing}. These results demonstrate that retrieval-aware training enables robust Boolean query generation, even when trained on a single corpus.

%% file: tables/result-base.tex
\begin{table*}[t]
	\footnotesize
	\centering
	\caption{Effectiveness of LLM-generated Boolean queries on the PubTemp set. \textbf{Bold} indicates the best result for each model within a setting; \underline{Underlined} indicates the overall best across all models. (O3 model has no N.R capability as reasoning was enabled by default using API.~\vspace{-9pt})}
	\label{tab:eval-results-pubmed}
		\begin{tabular*}{\textwidth}{@{\extracolsep{\fill}}lllllllllll@{}}
			\toprule
			%\multicolumn{3}{c}{} & \multicolumn{4}{c}{\textbf{Primary Metrics}} & \multicolumn{4}{c}{\textbf{Secondary Metrics}} \\
			\cmidrule(lr){4-7} \cmidrule(lr){8-11}
			Setting & Model & Prompt & Recall & F3 & \parbox{2.5em}{Recall\\>80\%} & \parbox{2.5em}{Recall\\>90\%} & Precision & \parbox{4em}{Avg\\Retrieved} & \parbox{2.5em}{Avg\\Regen} & \%Success \\
			\midrule
			\multirow{11}{*}{\rotatebox{90}{Zero-shot}} & GPT-4o & N.R & 0.3591 & \textbf{0.1758} & 12.40 & 7.10 & \underline{\textbf{0.1074}} & \textbf{251.32} & 1.15 & 99.60 \\
			& GPT-4o & R & 0.3937 & 0.1653 & 14.30 & 9.00 & 0.0913 & 326.20 & 1.09 & 99.80 \\
			& GPT-4o & R-con & \textbf{0.4387} & 0.1491 & \textbf{15.90} & \textbf{9.40} & 0.0642 & 440.45 & \underline{\textbf{1.04}} & \textbf{99.90} \\
			& GPT-4o & R-obj & 0.2530 & 0.0988 & 5.30 & 3.10 & 0.0742 & 359.00 & 1.20 & 99.80 \\
			\cmidrule{3-11}
			& O3 & R & \textbf{0.6868} & 0.2039 & \textbf{44.80} & \textbf{31.80} & 0.0611 & 551.54 & 1.69 & 98.20 \\
			& O3 & R-con & 0.6483 & \underline{\textbf{0.2106}} & 38.60 & 26.20 & \textbf{0.0690} & \textbf{499.48} & \textbf{1.19} & 99.70 \\
			& O3 & R-obj & 0.5153 & 0.1293 & 22.70 & 14.70 & 0.0510 & 603.10 & 1.21 & \underline{\textbf{100.00}} \\
			\cmidrule{3-11}
			& Qwen3-4B & N.R & 0.0098 & 0.0074 & 0.00 & 0.00 & 0.0175 & \underline{\textbf{64.72}} & 7.93 & 25.60 \\
			& Qwen3-4B & R & 0.0681 & 0.0429 & 0.70 & 0.60 & \textbf{0.0640} & 105.13 & 3.86 & 84.40 \\
			& Qwen3-4B & R-con & \textbf{0.3458} & \textbf{0.0824} & \textbf{11.60} & \textbf{6.50} & 0.0402 & 514.11 & \textbf{1.29} & \textbf{99.60} \\
			& Qwen3-4B & R-obj & 0.0676 & 0.0212 & 1.40 & 1.10 & 0.0233 & 306.53 & 2.89 & 93.80 \\
			\midrule
			\multirow{4}{*}{\rotatebox{90}{\shortstack{AutoBool \\ ($\alpha = 0.5$) \\ Weak R.O}}} & Qwen3-4B & N.R & \textbf{0.6791} & \textbf{0.1386} & \textbf{42.70} & \textbf{30.80} & \textbf{0.0392} & 677.31 & 1.12 & 98.80 \\
			& Qwen3-4B & R & 0.6112 & 0.1223 & 33.60 & 21.70 & 0.0345 & 678.80 & \underline{\textbf{1.04}} & 99.60 \\
			& Qwen3-4B & R-con & 0.5202 & 0.1052 & 22.80 & 14.30 & 0.0388 & \textbf{654.57} & 1.11 & \textbf{99.80} \\
			& Qwen3-4B & R-obj & 0.6495 & 0.0987 & 39.10 & 25.50 & 0.0263 & 743.89 & 1.10 & 99.40 \\
			\midrule
			\multirow{4}{*}{\rotatebox{90}{\shortstack{AutoBool \\ ($\alpha = 1$) \\ Mod R.O}}} & Qwen3-4B & N.R & \underline{\textbf{0.7036}} & 0.1195 & \underline{\textbf{47.10}} & \underline{\textbf{32.30}} & 0.0300 & 732.49 & 1.17 & 98.40 \\
			& Qwen3-4B & R & 0.5453 & \textbf{0.1346} & 27.90 & 16.50 & \textbf{0.0496} & \textbf{586.01} & 1.06 & 99.70 \\
			& Qwen3-4B & R-con & 0.5262 & 0.1066 & 26.30 & 16.80 & 0.0372 & 636.19 & 1.06 & \underline{\textbf{100.00}} \\
			& Qwen3-4B & R-obj & 0.6540 & 0.1094 & 39.70 & 26.70 & 0.0293 & 738.34 & \underline{\textbf{1.04}} & 99.90 \\
			\midrule
			\multirow{4}{*}{\rotatebox{90}{\shortstack{AutoBool \\ ($\alpha = 2$) \\ Heavy R.O}}} & Qwen3-4B & N.R & \textbf{0.6948} & 0.1209 & \textbf{45.30} & 30.80 & 0.0306 & 724.15 & 1.11 & 98.90 \\
			& Qwen3-4B & R & 0.5602 & \textbf{0.1344} & 30.00 & 18.40 & \textbf{0.0472} & \textbf{588.21} & 1.07 & 99.60 \\
			& Qwen3-4B & R-con & 0.5911 & 0.0887 & 31.40 & 20.40 & 0.0281 & 739.69 & 1.09 & \textbf{99.80} \\
			& Qwen3-4B & R-obj & 0.6878 & 0.1024 & 44.80 & \textbf{31.30} & 0.0245 & 773.01 & \textbf{1.06} & 99.70 \\
			\bottomrule
			
		\end{tabular*}
		~\vspace{-12pt}
\end{table*}

%% file: sections/ablations.tex
\section{Ablation Studies}
\label{sec:ablation}

%In our main ablation studies, we systematically evaluate how model size, training temperature, and backbone architecture affect AutoBool's query generation effectiveness. Unless otherwise specified, all models are trained on the same data using our reward function with \(\alpha = 1\). Appendix~\ref{appendix:retrieval_reward} provides comprehensive ablations on reward function design, including the necessity of logarithmic scaling, recall dependency, and precision weighting, as well as sensitivity analyses for the scaling factor M, smoothing constant s, and penalty values. The key finding was that the reasoning based prompt benefit from all of the different reward function component and was robust to different parameter settings. In contrast, the non-reasoning based prompt can be sensitive and therefore does necessitate careful consideration of parameter settings. 

In our main ablation studies, we systematically evaluate how model size, training temperature, and backbone architecture affect AutoBool's query generation effectiveness~\footnote{Unless otherwise specified, all models are trained on the same data using our reward function with \(\alpha = 1\).}.  Appendix~\ref{appendix:retrieval_reward} provides comprehensive ablations on reward function design, including logarithmic scaling, recall dependency, precision weighting, and sensitivity analyses for scaling factor M, smoothing constant s, and penalty values. The key finding is that 
while every component in the retrieval reward is important, reasoning-based prompts are more robust when hyperparameters vary, whereas no-reasoning prompts are more sensitive and require careful selection of hyperparameter values.

\subsection{Effect of Backbone Size}

To understand the impact of backbone model size on retrieval performance, we evaluate Qwen3 models at four parameter sizes (1.7B, 4B, 8B, and 14B), each trained with the same reinforcement learning setup. Figure~\ref{fig:model-size-comparison} shows that increasing model size leads to a consistent improvement in F\textsubscript{3}, reflecting better overall balance between recall and precision. However, this comes with a trade-off: recall and recall-threshold metrics (recall > 80\%, recall > 90\%) tend to decrease slightly as model size increases. This suggests that larger models may learn to generate more screening-efficient queries, retrieving fewer documents with higher precision, at the cost of missing some relevant studies.

These findings highlight the importance of aligning model scale with task priorities. For high-recall applications like systematic reviews, smaller models (e.g., 4B) may be preferable due to their stronger recall performance. In contrast, larger models (e.g., 8B or 14B) may be more appropriate when minimizing screening effort is critical and slight recall reductions are acceptable.

\subsection{Impact of Temperature}

\begin{figure*}[htbp]
	\centering
	\includegraphics[width=\textwidth]{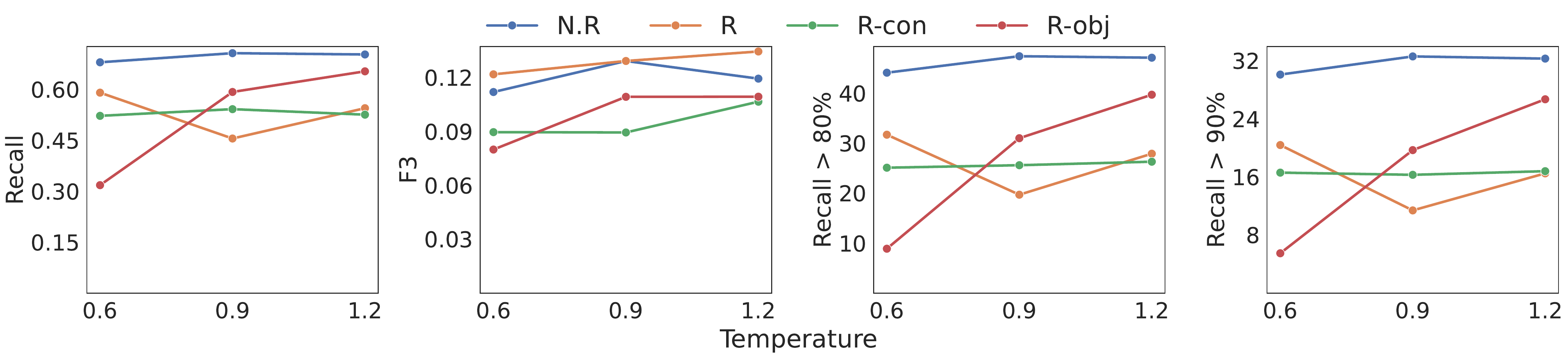}
	\caption{Effect of reinforcement learning training temperature value on the effectiveness of Boolean query generation across prompt types on the PubMed Temporal-Cutoff set,  all result based on Qwen3-4B Model.}
	\label{fig:temperature-comparison}
\end{figure*}

At training time, the generation temperature controls sampling diversity: higher values promote more varied outputs, while lower values encourage more deterministic decoding during training for the same topic. To assess its effect on retrieval effectiveness, we evaluate AutoBool models at three generation temperatures: 0.6, 0.9, and 1.2.
As shown in Figure~\ref{fig:temperature-comparison}, increasing temperature consistently improves all primary metrics. This indicates that more diverse generations help the model explore effective query formulations, resulting in broader coverage and higher-quality retrieval.

\subsection{Effect of Backbone Model }

\begin{table}[t]
	\centering
	\footnotesize
	\caption{Impact of Backbone model on the effectiveness of Boolean query generation across prompt types on the PubMed Temporal-Cutoff set.}
	\renewcommand{\arraystretch}{1.25}
	\label{tab:model_comparison}
		\begin{tabular*}{\columnwidth}{@{\extracolsep{\fill}}llcccc@{}}
			\toprule
			Model & Prompt & Recall & F3 & \parbox{2.5em}{Recall\\>80\%} & \parbox{2.5em}{Recall\\>90\%} \\
			\midrule
			\multirow{4}{*}{\rotatebox{90}{Qwen3-8B}} & N.R & 0.7116 & 0.1304 & 47.90 & 32.80 \\
			& R & 0.4716 & 0.1360 & 20.80 & 13.20 \\
			& R-con & 0.4928 & 0.1167 & 22.50 & 14.40 \\
			& R-obj & 0.5097 & \textbf{0.1376} & 19.50 & 12.30 \\
			\midrule
			\multirow{4}{*}{\rotatebox{90}{Llama3.1-8B}} & N.R & \textbf{0.7380} & 0.1035 & 53.00 & 38.00 \\
			& R & 0.7375 & 0.0999 & \textbf{54.20} & \textbf{39.70} \\
			& R-con & 0.7165 & 0.1006 & 48.60 & 34.00 \\
			& R-obj & 0.7291 & 0.1075 & 51.40 & 36.60 \\
			\bottomrule
		\end{tabular*}%

\end{table}

To examine how the backbone model impacts effectiveness, we compare two similarly sized models—Qwen3-8B and LLaMA3.1-8B—trained with identical reinforcement learning procedures. Table~\ref{tab:model_comparison} shows that LLaMA3.1-8B consistently outperforms Qwen3-8B across all primary recall metrics.
LLaMA3.1-8B obtains the highest recall across all prompt types, with strong results under \texttt{N.R} (0.7380 recall, 53.00\% recall > 80\%) and \texttt{R} (0.7375 recall, 54.20\% recall > 80\%). In contrast, Qwen3-8B yields higher F\textsubscript{3}, suggesting greater efficiency in reducing screening effort.

Consistent with findings on Qwen models, LLaMA3.1 is also more effective under the \texttt{No Reasoning} prompt, indicating that post-training, simpler prompting leads to more effective generation. Reasoning-based prompts tend to degrade recall performance, though they may still offer benefits in terms of robustness or interpretability.

Overall, these results highlight that backbone differences among decoder-only LLMs can significantly influence learning dynamics under reinforcement optimization, particularly in how recall and efficiency are balanced during query generation.

%% file: sections/conclusion.tex
\section{Conclusion}

We present \textbf{AutoBool}, a reinforcement learning framework for training language models to generate high-quality Boolean queries for systematic reviews. We also release PubTemp: a real-world dataset significantly larger than existing training and evaluation resources.

AutoBool provides scalable training by optimizing directly for retrieval effectiveness without requiring ground-truth queries. This overcomes a major limitation of supervised fine-tuning: the need for gold-standard Boolean queries, which are not available in large quantities for systematic reviews.
AutoBool substantially improves recall and robustness over zero-shot and few-shot methods using the same backbone, and matches or exceeds much larger models (GPT-4o and O3). Reinforcement learning enhances both retrieval effectiveness and generation stability: trained models consistently achieve higher recall, broader recall-threshold coverage, and stronger success rates. These improvements generalize to external test collections like CLEF and the Seed Collection, highlighting the transferability of retrieval-aware optimization.

Ablation studies reveal several key insights. Larger models improve F\textsubscript{3} but slightly reduce recall. Backbone choice matters: LLaMA is more effective than Qwen at the same scale. Higher decoding temperatures improve retrieval by increasing query diversity. Notably, No Reasoning prompts consistently yield the highest effectiveness after training but require carefully designed rewards, while reasoning-based prompts are more robust across different reward hyperparameters.

Together, these findings establish AutoBool as a scalable, flexible, and highly effective solution for automated Boolean query generation, balancing comprehensiveness, efficiency, and reliability for evidence-based search.

\section{Limitations}
While our findings demonstrate the effectiveness of AutoBool in generating high-recall Boolean queries with relatively small language models, several limitations remain.

First, we are currently limited to fine-tuning open-source models with moderate sizes (e.g., up to 14B parameters). Although AutoBool outperforms much larger commercial LLMs (e.g., GPT-4o, O3) in recall, prior work suggests that larger models may better support reasoning-based prompts and produce higher-quality queries. Due to GPU memory constraints, we are unable to fine-tune such models, limiting our ability to explore how retrieval-aware training scales with model size—particularly in prompt-sensitive settings.
Second, while LLaMA3.1 models achieved higher retrieval effectiveness in our experiments, their RL training was significantly less stable than Qwen3. We observed abrupt collapses in average reward during training that often did not recover, preventing reliable replication and leading us to focus on Qwen-based models for core experiments. We provide detailed analysis of this instability in Appendix~\ref{appendix:llama-instability}.
Third, as with most RL-fine-tuned LLMs, AutoBool exhibits stochastic behavior during training due to non-deterministic generation and moderate-temperature decoding. This can introduce minor variance across runs and affect reproducibility at the query level, though overall performance trends remain consistent.

Future work could improve training stability for architectures like LLaMA and investigate the root causes of instability. Access to larger and more robust open-source models may further enhance AutoBool’s effectiveness and generalizability.

%% file: sections/appendix.tex
\appendix

\section{Appendix}

\subsection{Prompt Template}

We design four prompt templates to investigate how different reasoning styles influence Boolean query generation performance:

\begin{itemize}
	\item \textbf{No Reasoning (N.R):} A simple prompt that directly asks the model to generate a Boolean query from a review topic without explanation. (See Table~\ref{fig:no-reasoning-prompt})
	\item \textbf{Free-text Reasoning (R):} Allows the model to reason freely before producing a final query, enabling unstructured decomposition. (See Table~\ref{fig:reasoning-prompt})
	\item \textbf{Conceptual Reasoning (R-con):} Guides the model to map the topic into structured elements like Population, Intervention, and Outcome before forming the query. (See Table~\ref{fig:prompt-conceptual})
	\item \textbf{Objective Reasoning (R-obj):} Encourages the model to extract explicit inclusion criteria and convert them into Boolean syntax. (See Table~\ref{fig:prompt-objective})
\end{itemize}

These templates serve as both zero-shot prompting strategies and scaffolds for reinforcement learning. Their comparative impact is analyzed throughout our experiments.

\label{appendix:prompts}

\input{tables/no-reasoning-prompt.tex}

\input{tables/free-reasoning-prompt.tex}

\input{tables/conceptual-reasoning-prompt.tex}

\input{tables/objective-reasoning-prompt.tex}

\subsection{Model Tuning}
\label{appendix:training-details}

\input{tables/tunning-parameter.tex}

\subsection{Ablation Result Comparison}

To evaluate AutoBool's generalization capability and validate its design choices, we conduct comprehensive ablation studies examining performance across multiple dimensions. We first assess generalization to existing datasets including CLEF TAR and Seed Collection (although the number of topics in these datasets are much less than the we published), demonstrating robustness across diverse systematic review domains. We then compare AutoBool against in-context learning baselines to isolate the contribution of reinforcement learning over prompt-based approaches. Finally, we analyze performance variations across individual topics to identify systematic patterns in query difficulty and failure modes. 

\subsubsection{Result on CLEF and Seed}

\begin{table*}[t]
	\centering
	\footnotesize
	\caption{Effectiveness of LLM-generated Boolean queries on the CLEF TAR set. \textbf{Bold} indicates the best result for each model within a setting; \underline{Underlined} indicates the overall best across all models. \textbf{Expert-Crafted} refers to results obtained by issuing the original Boolean queries from the dataset to PubMed. \textbf{Best~\citet{wang2025reassessing}-O1} denotes the O1 model using the P3 prompt from~\citet{wang2025reassessing}, which achieved the highest recall in that study.}
	\label{tab:eval-results-clef}
		\begin{tabular*}{\textwidth}{@{\extracolsep{\fill}}lllllllllll@{}}
			\toprule
			%\multicolumn{3}{c}{} & \multicolumn{4}{c}{\textbf{Primary Metrics}} & \multicolumn{4}{c}{\textbf{Secondary Metrics}} \\
			\cmidrule(lr){4-7} \cmidrule(lr){8-11}
			Setting & Model & Prompt & Recall & F3 & \parbox{2.5em}{Recall\\>80\%} & \parbox{2.5em}{Recall\\>90\%} & Precision & \parbox{4em}{Avg\\Retrieved} & \parbox{2.5em}{Avg\\Regen} & \%Success \\
			\midrule
			\multirow{12}{*}{\rotatebox{90}{Zero-shot}} & \multicolumn{2}{l}{Expert Crafted} & \underline{\textbf{0.8458}} & \textbf{0.0970} & \underline{\textbf{80.56}} & \underline{\textbf{79.17}} & \textbf{0.0206} & \textbf{14327.07} & / & \underline{\textbf{100.00}} \\
			& \multicolumn{2}{l}{Best~\citet{wang2025reassessing}-O1} & 0.6545 & 0.1966 & / & / & 0.1078 & / & / & / \\			\cmidrule{3-11}
			 & GPT-4o & N.R & 0.4258 & 0.2245 & 19.44 & 11.11 & \textbf{0.1275} & \textbf{389.74} & 1.01 & \underline{\textbf{100.00}} \\
			& GPT-4o & R & 0.4534 & 0.2160 & 22.22 & 16.67 & 0.1013 & 459.18 & 1.01 & \underline{\textbf{100.00}} \\
			& GPT-4o & R-con & \textbf{0.5283} & \textbf{0.2283} & \textbf{26.39} & \textbf{19.44} & 0.1080 & 535.54 & \underline{\textbf{1.00}} & \underline{\textbf{100.00}} \\
			& GPT-4o & R-obj & 0.3498 & 0.1449 & 16.67 & 9.72 & 0.1139 & 422.67 & 1.07 & \underline{\textbf{100.00}} \\
			\cmidrule{3-11}
			& O3 & R & \textbf{0.7454} & 0.3167 & \textbf{56.94} & \textbf{40.28} & 0.0879 & 663.10 & 1.58 & 98.61 \\
			& O3 & R-con & 0.7270 & \underline{\textbf{0.3196}} & 48.61 & 34.72 & \textbf{0.0928} & \textbf{627.35} & 1.18 & \underline{\textbf{100.00}} \\
			& O3 & R-obj & 0.5811 & 0.2157 & 25.00 & 13.89 & 0.0779 & 731.69 & \textbf{1.15} & \underline{\textbf{100.00}} \\
			\cmidrule{3-11}
			& Qwen3-4B & N.R & 0.0255 & 0.0235 & 0.00 & 0.00 & 0.0608 & \underline{\textbf{59.00}} & 6.39 & 41.67 \\
			& Qwen3-4B & R & 0.1756 & 0.1246 & 2.78 & 0.00 & \underline{\textbf{0.1389}} & 175.38 & 1.65 & 98.61 \\
			& Qwen3-4B & R-con & \textbf{0.5282} & \textbf{0.1579} & \textbf{33.33} & \textbf{22.22} & 0.0871 & 608.08 & \textbf{1.18} & \underline{\textbf{100.00}} \\
			& Qwen3-4B & R-obj & 0.0717 & 0.0422 & 1.39 & 1.39 & 0.0447 & 236.60 & 2.15 & 94.44 \\
			\midrule
			\multirow{4}{*}{\rotatebox{90}{\shortstack{AutoBool \\ ($\alpha = 0.5$) \\ Weak R.O}}} & Qwen3-4B & N.R & \textbf{0.7919} & \textbf{0.2497} & \textbf{65.28} & \textbf{41.67} & \textbf{0.0728} & 772.68 & 1.12 & 98.61 \\
			& Qwen3-4B & R & 0.6677 & 0.1773 & 43.06 & 31.94 & 0.0709 & 769.17 & \textbf{1.01} & \underline{\textbf{100.00}} \\
			& Qwen3-4B & R-con & 0.5729 & 0.1520 & 30.56 & 22.22 & 0.0548 & \textbf{722.76} & 1.39 & 97.22 \\
			& Qwen3-4B & R-obj & 0.7909 & 0.1704 & 62.50 & \textbf{41.67} & 0.0390 & 867.03 & \textbf{1.01} & \underline{\textbf{100.00}} \\
			\midrule
			\multirow{4}{*}{\rotatebox{90}{\shortstack{AutoBool \\ ($\alpha = 1$) \\ Mod R.O}}} & Qwen3-4B & N.R & \textbf{0.8387} & \textbf{0.2401} & \textbf{70.83} & \textbf{51.39} & 0.0499 & 818.31 & \underline{\textbf{1.00}} & \underline{\textbf{100.00}} \\
			& Qwen3-4B & R & 0.6090 & 0.2011 & 31.94 & 19.44 & \textbf{0.0782} & \textbf{696.92} & 1.06 & \underline{\textbf{100.00}} \\
			& Qwen3-4B & R-con & 0.5885 & 0.1684 & 34.72 & 20.83 & 0.0672 & 697.35 & 1.03 & \underline{\textbf{100.00}} \\
			& Qwen3-4B & R-obj & 0.7619 & 0.2035 & 55.56 & 43.06 & 0.0433 & 880.74 & \underline{\textbf{1.00}} & \underline{\textbf{100.00}} \\
			\midrule
			\multirow{4}{*}{\rotatebox{90}{\shortstack{AutoBool \\ ($\alpha = 2$) \\ Heavy R.O}}} & Qwen3-4B & N.R & \textbf{0.8239} & \textbf{0.2193} & \textbf{73.61} & \textbf{51.39} & 0.0538 & 830.08 & \underline{\textbf{1.00}} & \underline{\textbf{100.00}} \\
			& Qwen3-4B & R & 0.5723 & 0.2054 & 31.94 & 15.28 & \textbf{0.0737} & \textbf{698.60} & \underline{\textbf{1.00}} & \underline{\textbf{100.00}} \\
			& Qwen3-4B & R-con & 0.6571 & 0.1348 & 41.67 & 29.17 & 0.0356 & 869.71 & 1.04 & \underline{\textbf{100.00}} \\
			& Qwen3-4B & R-obj & 0.8177 & 0.1995 & 65.28 & \textbf{51.39} & 0.0406 & 904.31 & \underline{\textbf{1.00}} & \underline{\textbf{100.00}} \\
			\bottomrule
		\end{tabular*}
\end{table*}

\begin{table*}[t]
	\centering
	\footnotesize
	\caption{Effectiveness of LLM-generated Boolean queries on Seed Collection.. \textbf{Bold} indicates the best result for each model within a setting; \underline{Underlined} indicates the overall best across all models. \textbf{Expert-Crafted} refers to results obtained by issuing the original Boolean queries from the dataset to PubMed. \textbf{Best~\citet{wang2025reassessing}-O1} denotes the O1 model using the P3 prompt from~\citet{wang2025reassessing}, which achieved the highest recall in that study}
	\label{tab:eval-results-seed}
		\begin{tabular*}{\textwidth}{@{\extracolsep{\fill}}lllllllllll@{}}
			\toprule
			%\multicolumn{3}{c}{} & \multicolumn{4}{c}{\textbf{Primary Metrics}} & \multicolumn{4}{c}{\textbf{Secondary Metrics}} \\
			\cmidrule(lr){4-7} \cmidrule(lr){8-11}
			Setting & Model & Prompt & Recall & F3 & \parbox{2.5em}{Recall\\>80\%} & \parbox{2.5em}{Recall\\>90\%} & Precision & \parbox{4em}{Avg\\Retrieved} & \parbox{2.5em}{Avg\\Regen} & \%Success \\
			\midrule
			
			\multirow{12}{*}{\rotatebox{90}{Zero-shot}} &  \multicolumn{2}{l}{Expert Crafted} & \underline{\textbf{0.7241}} & \underline{\textbf{0.1869}} & \underline{\textbf{57.50}} & \textbf{25.00} & 0.0341 & 1416.55 & / & \underline{\textbf{100.00}} \\
			& \multicolumn{2}{l}{Best~\citet{wang2025reassessing}-O1} & 0.5786 & 0.0852 & / & / & 0.0523 & / & / & / \\
			\cmidrule{3-11}
			& GPT-4o & N.R & 0.3106 & 0.1073 & 2.50 & \textbf{2.50} & 0.0694 & \textbf{369.30} & \underline{\textbf{1.00}} & \underline{\textbf{100.00}} \\
			& GPT-4o & R & 0.3330 & 0.1076 & 5.00 & \textbf{2.50} & 0.0317 & 426.77 & 1.05 & \underline{\textbf{100.00}} \\
			& GPT-4o & R-con & \textbf{0.3936} & \textbf{0.1262} & 2.50 & 0.00 & 0.0382 & 559.98 & \underline{\textbf{1.00}} & \underline{\textbf{100.00}} \\
			& GPT-4o & R-obj & 0.2647 & 0.0847 & \textbf{7.50} & \textbf{2.50} & \underline{\textbf{0.0882}} & 478.00 & 1.25 & \underline{\textbf{100.00}} \\
			\cmidrule{3-11}
			& O3 & R & \textbf{0.7027} & 0.1142 & \textbf{47.50} & \textbf{25.00} & 0.0174 & 733.10 & 1.48 & \underline{\textbf{100.00}} \\
			& O3 & R-con & 0.6482 & 0.1336 & 30.00 & 15.00 & 0.0235 & 659.42 & 1.35 & \underline{\textbf{100.00}} \\
			& O3 & R-obj & 0.5411 & \textbf{0.1418} & 22.50 & 12.50 & \textbf{0.0424} & \textbf{653.65} & \textbf{1.12} & \underline{\textbf{100.00}} \\
			\cmidrule{3-11}
			& Qwen3-4B & N.R & 0.0003 & 0.0000 & 0.00 & 0.00 & 0.0000 & \underline{\textbf{124.00}} & 8.10 & 25.00 \\
			& Qwen3-4B & R & 0.0306 & 0.0144 & 0.00 & 0.00 & \textbf{0.0327} & 170.43 & 4.35 & 75.00 \\
			& Qwen3-4B & R-con & \textbf{0.3768} & \textbf{0.0567} & \textbf{10.00} & \textbf{10.00} & 0.0193 & 596.70 & \textbf{1.23} & \underline{\textbf{100.00}} \\
			& Qwen3-4B & R-obj & 0.0384 & 0.0113 & 0.00 & 0.00 & 0.0275 & 194.13 & 2.58 & 97.50 \\
			\midrule
			\multirow{4}{*}{\rotatebox{90}{\shortstack{AutoBool \\ ($\alpha = 0.5$) \\ Weak R.O}}} & Qwen3-4B & N.R & 0.5820 & 0.0855 & \textbf{37.50} & 20.00 & 0.0126 & 803.08 & 1.23 & 97.50 \\
			& Qwen3-4B & R & 0.5445 & \textbf{0.0899} & 27.50 & 25.00 & 0.0147 & 726.08 & \textbf{1.02} & \underline{\textbf{100.00}} \\
			& Qwen3-4B & R-con & 0.4701 & 0.0873 & 17.50 & 15.00 & \textbf{0.0198} & \textbf{651.63} & 1.48 & 95.00 \\
			& Qwen3-4B & R-obj & \textbf{0.6419} & 0.0820 & \textbf{37.50} & \textbf{27.50} & 0.0170 & 817.83 & 1.15 & \underline{\textbf{100.00}} \\
			\midrule
			\multirow{4}{*}{\rotatebox{90}{\shortstack{AutoBool \\ ($\alpha = 1$) \\ Mod R.O}}} & Qwen3-4B & N.R & \textbf{0.6828} & 0.0943 & \textbf{47.50} & \underline{\textbf{35.00}} & 0.0136 & 827.00 & 1.23 & 97.50 \\
			& Qwen3-4B & R & 0.5301 & 0.0668 & 30.00 & 17.50 & 0.0109 & \textbf{735.50} & \underline{\textbf{1.00}} & \underline{\textbf{100.00}} \\
			& Qwen3-4B & R-con & 0.5307 & \textbf{0.1051} & 30.00 & 12.50 & \textbf{0.0226} & 736.30 & 1.15 & \underline{\textbf{100.00}} \\
			& Qwen3-4B & R-obj & 0.5890 & 0.0642 & 35.00 & 27.50 & 0.0097 & 831.35 & 1.05 & \underline{\textbf{100.00}} \\
			\midrule
			\multirow{4}{*}{\rotatebox{90}{\shortstack{AutoBool \\ ($\alpha = 2$) \\ Heavy R.O}}} & Qwen3-4B & N.R & 0.6539 & 0.0749 & 35.00 & 25.00 & 0.0099 & 855.03 & 1.25 & 97.50 \\
			& Qwen3-4B & R & 0.5963 & \textbf{0.1071} & 32.50 & 20.00 & \textbf{0.0215} & \textbf{730.12} & 1.05 & \underline{\textbf{100.00}} \\
			& Qwen3-4B & R-con & 0.5066 & 0.0528 & 20.00 & 17.50 & 0.0123 & 786.40 & \underline{\textbf{1.00}} & \underline{\textbf{100.00}} \\
			& Qwen3-4B & R-obj & \textbf{0.6804} & 0.0672 & \textbf{50.00} & \underline{\textbf{35.00}} & 0.0089 & 896.00 & 1.23 & 97.50 \\
			\bottomrule
		\end{tabular*}
\end{table*}

We report detailed evaluation results for AutoBool on two widely used external benchmarks for Boolean query generation: CLEF TAR (Table~\ref{tab:eval-results-clef}) and the Seed Collection (Table~\ref{tab:eval-results-seed}). Both tables compare AutoBool against zero-shot prompting baselines using the same underlying model (Qwen3-4B), commercial LLMs (GPT-4o, O3), and manually written expert queries when available. These benchmarks provide insights into AutoBool's generalization ability when deployed beyond its training distribution.

\paragraph{CLEF TAR.} AutoBool demonstrates strong generalization across all primary metrics. It substantially improves recall over zero-shot baselines and also outperforms commercial LLMs such as GPT-4o and O3 on recall and high-recall coverage (Recall > 80\%, Recall > 90\%). Notably, when using the \texttt{No Reasoning} prompt and \(\alpha = 1\), AutoBool achieves recall that is within 1\% of expert-written queries, while retrieving 17× fewer documents. This trade-off leads to improved F\textsubscript{3} and precision, indicating not only strong comprehensiveness but also a meaningful reduction in screening burden. AutoBool also surpasses the performance of the O1 model used in~\citet{wang2025reassessing}, highlighting the advantage of retrieval-aware training via reinforcement learning over static prompting-based generation.

\paragraph{Seed Collection.} Similar results are observed on the Seed Collection benchmark. While AutoBool does not fully match the performance of expert-authored queries in recall or F\textsubscript{3}, it consistently outperforms all zero-shot prompting baselines and the O1 model from~\citet{wang2025reassessing}. The model demonstrates strong recall-oriented behavior while keeping the number of retrieved documents at a practical level, maintaining its advantage in terms of screening efficiency. The success rate remains close to 100\%, indicating stable formatting and reliable generation.

These results confirm that AutoBool generalizes effectively across different domains and collections, even though it was trained solely on PubMed data. The performance gains on CLEF and Seed further reinforce the value of reinforcement learning for robust and transferable Boolean query generation.

\subsubsection{Comparison to In Context Learning}
\label{appendix:icl}
\begin{table}[t!]
	\centering
	\small
	\caption{Performance comparison of Zero-shot, In-context Learning (ICL), and the Autobool for the Qwen3-4B model. Best performance for each metric within a prompt group is highlighted in bold.}
	\label{tab:method_comparison_qwen3-4b}

	\begin{tabular}{@{}llcccc@{}}
		\toprule
		Prompt & Method & Recall & F3 & \parbox{2.5em}{Recall\\>80\%}  & \parbox{2.5em}{Recall\\>90\%}  \\
		\midrule
		\multirow{5}{*}{N.R} & Zero-shot & 0.0098 & 0.0074 & 0.00 & 0.00 \\
		& ICL (1-shot) & 0.1011 & 0.0334 & 3.80 & 2.80 \\
		& ICL (3-shot) & 0.0899 & 0.0337 & 2.50 & 1.20 \\
		& ICL (5-shot) & 0.0848 & 0.0347 & 1.50 & 0.70 \\
		& Autobool & \textbf{0.7036} & \textbf{0.1195} & \textbf{47.10} & \textbf{32.30} \\
		\midrule
		\multirow{5}{*}{R} & Zero-shot & 0.0681 & 0.0429 & 0.70 & 0.60 \\
		& ICL (1-shot) & 0.1142 & 0.0410 & 2.80 & 2.00 \\
		& ICL (3-shot) & 0.1052 & 0.0419 & 3.00 & 2.20 \\
		& ICL (5-shot) & 0.1156 & 0.0442 & 2.80 & 1.80 \\
		& Autobool & \textbf{0.5453} & \textbf{0.1346} & \textbf{27.90} & \textbf{16.50} \\
		\midrule
		\multirow{5}{*}{R-con} & Zero-shot & 0.3458 & 0.0824 & 11.60 & 6.50 \\
		& ICL (1-shot) & 0.3341 & 0.0692 & 12.50 & 7.40 \\
		& ICL (3-shot) & 0.3052 & 0.0670 & 9.20 & 5.40 \\
		& ICL (5-shot) & 0.3094 & 0.0694 & 9.80 & 6.00 \\
		& Autobool & \textbf{0.5262} & \textbf{0.1066} & \textbf{26.30} & \textbf{16.80} \\
		\midrule
		\multirow{5}{*}{R-obj} & Zero-shot & 0.0676 & 0.0212 & 1.40 & 1.10 \\
		& ICL (1-shot) & 0.0812 & 0.0228 & 1.80 & 1.20 \\
		& ICL (3-shot) & 0.0969 & 0.0316 & 1.80 & 1.10 \\
		& ICL (5-shot) & 0.0799 & 0.0304 & 1.40 & 1.10 \\
		& Autobool & \textbf{0.6540} & \textbf{0.1094} & \textbf{39.70} & \textbf{26.70} \\
		\bottomrule

	\end{tabular}

\end{table}

To assess the effectiveness of our reinforcement learning approach against prompt-based alternatives, we compare AutoBool with zero-shot and in-context learning (ICL) baselines using the Qwen3-4B backbone model. For ICL, we provide exemplar systematic review topics with their corresponding expert-crafted Boolean queries from the CLEF collection~\cite{kanoulas2018clef}. For one-shot, we used topic CD010438, following~\citet{wang2025reassessing, wang2023can}.\footnote{Prior work has shown that using one high-quality example is more effective than using a related but lower-quality example.} For three-shot, we additionally selected topics CD007427 and CD009944; for five-shot, we added CD011602 and CD010213. All selected topics have comprehensive expert-formulated queries that satisfy our prompt requirements.

For reasoning-based prompts, to ensure the example reasoning patterns align with the inference distribution, we used the same inference prompt structure but provided the expert Boolean query and asked Qwen3-4B to generate the corresponding reasoning that would lead to that target query.

Table~\ref{tab:method_comparison_qwen3-4b} reveals that while ICL improves over zero-shot baselines in most cases, it demonstrates no scaling benefits with additional examples. For no-reasoning prompts, ICL improves recall from 0.01 (zero-shot) to 0.101 (1-shot), but then degrades to 0.085 (5-shot). The R prompt shows similar patterns with modest gains over zero-shot (0.068) reaching 0.114–0.116 with examples. Notably, R-con is the exception where ICL actually harms performance: declining from 0.346 (zero-shot) to 0.334 (1-shot) and further to 0.309 (5-shot). This absence of scaling or even negative scaling contrasts with typical NLP tasks where few-shot learning provides consistent improvements, suggesting that Boolean query generation requires learning formulation \emph{strategies} rather than pattern matching from examples.

Despite ICL's improvements over zero-shot, AutoBool substantially outperforms all ICL configurations across every prompt type and metric. For no-reasoning prompts, AutoBool achieves 0.704 recall—7.0× better than the best ICL result (1-shot: 0.101) and 47.1\% of topics exceeding 80\% recall versus only 3.8\% for best ICL. The R-obj prompt demonstrates similar dramatic improvement: 0.654 recall versus 0.097 for best ICL (3-shot), a 6.7× gain. Even for R-con where zero-shot outperforms ICL, AutoBool achieves 0.526 recall, improving upon both zero-shot (0.346) and ICL variants.

These results demonstrate that while examples can provide modest gains over zero-shot (except for R-con), they fail to capture the complexity of balancing recall-precision trade-offs, applying proper Boolean operators, and selecting appropriate field specifications. The flat or declining ICL performance with additional examples confirms that reinforcement learning with retrieval-based feedback is necessary for this task, enabling adaptive strategy learning through direct optimization rather than static pattern matching.

\subsubsection{Topics Analysis}
To understand performance variations across different systematic review topics, we conduct a comprehensive analysis of recall distributions and identify characteristic patterns distinguishing high-performing from low-performing queries. We partition our test data into three equal groups based on recall performance: top third (highest recall), middle third, and bottom third (lowest recall). We then analyze term frequency statistics within each category to identify systematic patterns in query characteristics.

\paragraph{Performance distribution demonstrates strong generalization.} Our models exhibit top-skewed recall distributions, indicating robust performance across most topics. For no-reasoning prompts, the top third of topics achieve approximately 0.90 recall, while reasoning-based prompts reach approximately 0.74 recall in their top third. This distribution pattern demonstrates that the models successfully generalize across diverse systematic review domains rather than memorizing specific topic patterns from training data.

\paragraph{Topic difficulty correlates with terminology specificity and complexity.} Through term frequency analysis comparing high-recall versus low-recall topics, we identify systematic patterns in query difficulty. High-recall topics typically involve specific medical procedures or distinct conditions with well-established terminology, such as ``portoenterostomy,'' ``keratoconus,'' and ``zuranolone.'' These terms have clear, unambiguous meanings in medical literature, facilitating precise Boolean query construction.

In contrast, low-recall topics exhibit three characteristic patterns. First, procedural or technical terms like ``root canals'' and ``ultrasonic clamping'' require precise field specifications that models struggle to generate correctly. Second, multi-component medical concepts such as ``platinumsensitiveresistant'' present challenges in formulating comprehensive queries that capture all relevant term variants. Third, broad conceptual terms like ``validation'' and ``algorithm'' lack the specificity needed for effective Boolean retrieval, leading to either overly broad or overly narrow queries.

\paragraph{Prompt-specific failure patterns reveal architectural limitations.} No-reasoning prompts demonstrate particular difficulty with topics requiring nuanced terminology distinctions. For example, dental procedures involve subtle semantic differences that significantly impact retrieval effectiveness, yet the model struggles to capture these distinctions without explicit reasoning guidance. Reasoning-based prompts face different challenges: they struggle with compound medical terms and emerging biomedical technologies such as ``multiomics'' and ``gwas.'' In these cases, the reasoning steps may introduce additional complexity or verbosity without corresponding improvements in query precision.

These findings suggest three possible improvement directions. First, enhanced handling of compound terms through specialized tokenization or term expansion strategies could better capture multi-component medical concepts. Second, improved field-specific query formulation through domain-aware templates could help models generate appropriate field tags for procedural topics. Third, specialized training strategies for emerging biomedical concepts could address terms that lack established Boolean query patterns in existing literature. The systematic nature of these failure modes indicates that architectural modifications or training data augmentation could address specific weaknesses rather than requiring fundamental redesign of the approach.

\subsection{Ablations on Retrieval Reward}
\label{appendix:retrieval_reward}

In this section, we systematically examine the design choices underlying our retrieval reward function. The reward function balances multiple objectives: recall prioritization, precision awareness, and poor-performing query awareness through recall-dependent precision weighting, logarithmic scaling, and penalty mechanisms. To validate these design decisions, we conduct comprehensive ablation studies that isolate the contribution of each component. These ablations reveal the empirical foundations of our reward design and provide practical guidance for hyperparameter selection in similar reinforcement learning applications for information retrieval tasks.

\subsubsection{Impact of Retrieval Reward Modules}
\input{tables/impact-retrieval-reward.tex}

To understand the contribution of each component in our reward function design, we conduct systematic ablation studies by removing key components: logarithmic scaling (W/O Scaling), recall dependency on precision reward (W/O Dependency), precision component (W/O Precision). These experiments demonstrate the necessity of each component for achieving balanced performance across different prompt types using the Qwen3-4B model on the PubTemp dataset.

To clarify the impact of each component, we define the modified reward functions used in our ablations:
\begin{itemize}
	\item \textbf{W/O Scaling:}
	\[
	F_{\text{no-log}}(r, p) = M \cdot r + M \cdot r^\alpha \cdot p
	\]
	\item \textbf{W/O Dependency:}
	\[
	F_{\text{no-dep}}(r, p) = M \cdot r + M \cdot p
	\]
	\item \textbf{W/O Precision:}
	\[
	F_{\text{no-prec}}(r, p) = M \cdot r
	\]
\end{itemize}

Table~\ref{tab:eval-results-reward-components} demonstrates the impact of different reward function components across all prompt types. The results reveal three critical insights about the necessity of each component for systematic review Boolean query generation.

\paragraph{Logarithmic scaling prevents catastrophic F3 degradation while maintaining recall quality.} The W/O Scaling variant shows dramatically different behavior patterns across prompt types. For no-reasoning prompts (N.R), removing logarithmic scaling yields higher raw recall (76.3\% vs 70.4\%) and improved high-recall coverage (56.6\% vs 47.1\% for recall $>80\%$), but causes severe F3 collapse from 0.1195 to 0.0429 (-64.1\%). This pattern indicates that without logarithmic scaling, models optimize for recall at the expense of precision control, creating queries that retrieve excessive irrelevant documents. Conversely, reasoning-based prompts show more stable behavior, with W/O Scaling causing modest F3 degradation (-7.6\% to -25.1\%) while maintaining reasonable recall performance. This suggests that explicit reasoning steps provide additional structure that partially compensates for the loss of logarithmic precision scaling.

\paragraph{Recall dependency critically balances recall-precision trade-offs across prompt architectures.} The impact of removing recall dependency (W/O Dependency) varies significantly by prompt type, revealing fundamental differences in how implicit versus explicit reasoning approaches handle reward optimization. For no-reasoning prompts, removing the $r^\alpha$ weighting causes substantial recall degradation (-35.2\%) while maintaining F3 performance, indicating premature precision optimization that sacrifices coverage. However, reasoning-based prompts exhibit opposite behavior: W/O Dependency actually improves recall performance (+16.8\% for R, +4.2\% for R-con, +2.0\% for R-obj) while showing modest F3 degradation. This counterintuitive result suggests that structured reasoning approaches benefit from independent precision optimization, as the explicit reasoning steps provide sufficient recall guidance without requiring mathematical dependency encoding.

\paragraph{Precision component removal consistently degrades balanced performance across all architectures.} The W/O Precision variant, which reduces the reward to pure recall optimization ($M \cdot r$), shows consistent negative impacts across all prompt types. F3 scores drop substantially (-28.8\% to -12.0\% for N.R and R respectively), while recall performance also degrades moderately (-8.8\% to -10.5\%). This pattern demonstrates that even recall-oriented systematic review applications require precision awareness to generate practically useful queries. Pure recall optimization leads to queries that retrieve excessive document volumes, creating unrealistic screening burdens that compromise the systematic review workflow.

The differential sensitivity patterns across prompt types reveal important architectural insights for reward design in reasoning-heavy applications. No-reasoning prompts require carefully balanced reward components with logarithmic scaling and recall dependency to achieve optimal performance, while reasoning-based approaches show greater robustness to individual component variations due to their structured intermediate outputs.

\subsubsection{Why not using F3 for retrieval reward}

\input{tables/impact-f3.tex}

While we design this complex reward, one question remains: what if we use a simpler reward that prioritises recall, like F3? To address this, we compare our decomposed reward design against F3 optimization using:
\[
F_{\text{F3}}(r, p)  = M \cdot \frac{(1 + \beta^2) \cdot r \cdot p}{\beta^2 \cdot r + p}
\]

where $\beta = 3$ for F3 weighting.

Table~\ref{tab:eval-results-reward-f3} shows that while F3 optimization achieves higher F3 scores (+11.2\% to +72.8\%), it causes severe recall degradation (-15.7\% to -48.0\%). Most critically, high-recall performance collapses: for recall $>80\%$, performance drops from 47.1\% to 12.0\% topics for N.R and from 39.7\% to 12.5\% for R-obj. These recall levels (0.37-0.44) fall far below the 0.80-0.90 thresholds required for systematic review quality standards, making F3-optimized queries practically unusable despite superior F3 scores.

\subsubsection{Impact of Scalling factor M}

\input{tables/impact-m.tex}

The scaling factor $M$ controls the magnitude of retrieval reward signals relative to the fixed format and validity rewards ($\pm 10$ each). Since $R_{\text{total}} = R_{\text{format}} + R_{\text{validity}} + R_{\text{retrieval}}$, the value of $M$ determines whether the model prioritizes syntactic correctness (low $M$) or retrieval effectiveness (high $M$). We examine $M$ values of 5, 10, and 20 to understand how this balance affects learning dynamics.

Table~\ref{tab:eval-results-m-values} reveals distinct sensitivity patterns across prompt architectures. The no-reasoning prompt (N.R) demonstrates high sensitivity to $M$ values, with optimal performance at $M=10$ achieving 0.7036 recall and 47.1\% high-recall coverage. Both lower ($M=5$) and higher ($M=20$) scaling values lead to substantial performance degradation, with recall dropping to 0.4368 and 0.5524 respectively. This sensitivity reflects the prompt's reliance on implicit learning without structured intermediate steps, making it dependent on precise reward signal calibration for effective policy optimization.

In contrast, reasoning-based prompts exhibit greater stability across different $M$ values. For structured reasoning approaches (R, R-con, R-obj), performance variations remain modest, with some prompts even achieving peak performance at higher scaling factors ($M=20$ for R and R-obj). This robustness suggests that explicit reasoning steps provide additional training signals that compensate for suboptimal reward scaling, making these approaches more resilient to hyperparameter variations. The structured intermediate outputs effectively regularize the learning process, reducing dependence on the primary reward scaling parameter.

\subsubsection{Impact of Smoothing constant s}
\input{tables/impact-s.tex}

The smoothing constant $s$ in the logarithmic scaling function $\log_{1+s}(1 + s \cdot p)$ controls the curvature of precision rewards, determining how quickly precision improvements translate into reward increases. Lower $s$ values create steeper gradients at low precision, while higher values provide more gradual scaling across the precision range.

Table~\ref{tab:eval-results-scale-values} demonstrates that $s=100$ provides optimal performance for the no-reasoning prompt, achieving 70.4\% recall and 47.1\% high-recall coverage, substantially outperforming both $s=10$ (38.9\% recall) and $s=1000$ (48.1\% recall). This indicates that moderate logarithmic curvature effectively balances gradient sensitivity at low precision values with stable training dynamics for implicit learning approaches.

Reasoning-based prompts show more varied optimal values across the range. The R prompt performs best at $s=10$ for recall metrics, while R-con and R-obj achieve peak performance at $s=1000$ and $s=10$ respectively. This variability suggests that structured reasoning approaches are less sensitive to precision reward curvature, as their explicit intermediate steps provide additional learning signals that compensate for suboptimal logarithmic scaling parameters.

\subsubsection{Impact of Retrieval Reward Penalty}

\input{tables/impact-retrieval-panelty.tex}

Penalty parameters guide model behavior by providing negative signals for undesirable outcomes in the retrieval reward function. We examine two key penalties in $R_{\text{retrieval}}(r, p, |D|)$: the empty result penalty (applied when $|D| = 0$) and the zero relevant results penalty (applied when $r = 0 \land p = 0$).

\paragraph{Empty result penalty maintains training stability through symmetric incentives.} Table~\ref{tab:eval-results-min-reward} shows that the penalty value of -20 achieves optimal performance across most prompt types. For no-reasoning prompts, this penalty yields 70.4\% recall and 47.1\% high-recall coverage, substantially outperforming both weaker (-10: 52.7\% recall) and stronger (-30: 46.2\% recall) penalties. The -20 value maintains intentional symmetry with the maximum retrieval reward (+20), creating balanced incentives between achieving good retrieval and avoiding complete failure states. Insufficient penalties (-10) fail to discourage empty queries adequately, while excessive penalties (-30) can cause training instability as models become overly cautious. Reasoning-based prompts show similar patterns with more modest performance variations, suggesting their structured outputs provide additional stability that partially compensates for suboptimal penalty values.

\paragraph{Zero relevant results penalty provides graduated failure signals.} Table~\ref{tab:eval-results-off-topic-penalty} demonstrates that the -5 penalty for queries retrieving documents but finding zero relevant results achieves consistently strong performance. This penalty represents less severe failure than empty results, as these queries demonstrate syntactic validity and sufficient topic relevance to retrieve documents despite poor relevance matching. The graduated penalty structure guides models toward incremental improvements rather than binary success/failure optimization. Without this penalty (0), performance degrades substantially: no-reasoning recall drops from 70.4\% to 41.8\%, while excessive penalties (-10) show modest degradation (50.7\% recall). The -5 value effectively balances encouraging document retrieval while maintaining pressure for relevance.

\subsection{Llama Model Training instability}

\label{appendix:llama-instability}
While LLaMA3.1-8B achieved the highest recall performance in our Boolean query generation experiments (Table~\ref{tab:model_comparison}), its RL training exhibited significant instability that prevented reliable deployment. 
%Training instability observed with LLaMA 3.1 models provides important insights into architecture-specific challenges for reinforcement learning in reasoning-heavy applications. 
Through systematic analysis of training logs and failure patterns, we identify three primary factors contributing to instability in the context of Boolean query generation for systematic reviews.

\paragraph{Architecture-prompt interaction effects.} Instability correlates strongly with prompt type complexity. LLaMA 3.1 models experience reward collapses primarily with reasoning-based prompts, following a clear severity hierarchy: free-text reasoning exhibits highest instability, structured reasoning shows moderate issues, and no-reasoning prompts remain relatively stable. This pattern contrasts sharply with Qwen3 models, which maintain stable training dynamics across all prompt types. We hypothesize this difference stems from Qwen3's pretraining on reasoning-specific tasks and long-context data, capabilities that LLaMA 3.1 lacks in comparable depth, making it more vulnerable to gradient instabilities when generating extended reasoning traces during RL optimization.

\paragraph{Generation length correlates with instability patterns.} Training failures increase systematically with output length requirements: free-text reasoning ($\sim$2000 tokens) demonstrates highest instability, structured reasoning ($\sim$1500 tokens) exhibits moderate issues, and no-reasoning ($\sim$100 tokens) remains stable. During collapse events, LLaMA 3.1 frequently generates repetitive content that exceeds the 3072 token limit, entering degenerate states that resist recovery through continued training. This suggests that the model's attention mechanisms may struggle to maintain coherent long-form generation under the gradient pressures of reinforcement learning optimization.

\paragraph{Pretraining differences explain stability variations.} Qwen3's technical documentation indicates specific reasoning-based and long-context training during pretraining phases, capabilities that LLaMA 3.1 lacks in comparable depth~\cite{yang2025qwen3}. This architectural preparation appears crucial for maintaining stability during RL optimization with reasoning-heavy prompts, providing the foundational capabilities necessary for stable policy learning in complex generation tasks.

These findings offer practical guidance for practitioners applying RL fine-tuning to Boolean query generation. The observed instability patterns indicate that GRPO's interaction with LLaMA's architecture becomes problematic when generating long, reasoning-heavy Boolean queries, particularly for model backbones without reasoning-specific pretraining. For Boolean query formulation tasks requiring extended reasoning traces, practitioners should either: (1) use models with reasoning-specific pretraining, or (2) employ no-reasoning prompts with LLaMA models to maintain training stability. Future work should investigate whether similar stability issues appear in other structured query generation tasks and develop architecture-specific training strategies for complex retrieval applications.

\subsection{Example Case}
\label{appendix:example}

To illustrate behavioral differences across prompt types, we present an example systematic review topic 39707254: ``The asymptomatic tuberculosis proportion among the elderly population,'' which includes 28 studies as relevant studies. Figures~\ref{fig:topic_39707254_no_reasoning}--\ref{fig:topic_39707254_objective_reasoning} show the generated queries and performance metrics.

\paragraph{Generation patterns reveal distinct query construction strategies.} The no-reasoning prompt generates exhaustive term expansion through comprehensive OR-combinations across multiple semantic dimensions: tuberculosis variants (``TB,'' ``Mycobacterium tuberculosis''), asymptomatic concepts (``undiagnosed,'' ``latent''), elderly terms (``geriatric,'' ``aged,'' ``older adult''), and epidemiological terms (``prevalence,'' ``incidence,'' ``screening''). Free-text reasoning exhibits explicit deliberation in the \texttt{<think>} section, considering MeSH terms and field specifications, but produces notably more conservative queries that focus on core concepts. Conceptual reasoning systematically decomposes topics into constituent concepts (``asymptomatic tuberculosis,'' ``elderly population'') and generates extensive synonym lists for each before combining them. Objective reasoning simulates relevant article structures and incorporates study design terms (``cohort study,'' ``cross-sectional study'') alongside medical concepts, demonstrating domain-aware query construction.

\paragraph{Performance metrics reflect the recall-precision trade-offs of each strategy.} No-reasoning and conceptual reasoning both achieve perfect recall (1.0) through their exhaustive term coverage, but suffer from extremely low precision (0.0001 and 0.0000 respectively), retrieving over 270,000 documents. Free-text reasoning's conservative approach produces moderate recall (0.333) with improved precision (0.0038), though at the cost of missing two-thirds of relevant studies. Objective reasoning represents an intermediate strategy, yielding strong recall (0.667) with modest precision improvement (0.0002). These results demonstrate how different reasoning architectures fundamentally alter the query construction process and resulting retrieval characteristics.

\clearpage
\include{tables/example.tex}

\include{tables/example-b.tex}

%% file: tables/no-reasoning-prompt.tex
\begin{figure*}[ht!]
	\centering
	\small
	\begin{PromptBox}[System Message]
		You are an expert systematic review information specialist.
		
		You are tasked to formulate a systematic review Boolean query in response to a research topic. The final Boolean query must be enclosed within <answer> </answer> tags. Do not include any explanation or reasoning.
	\end{PromptBox}
	
	\vspace{1em}
	
	\begin{PromptBox}[User Message]
		You are given a systematic review research topic, with the topic title "{topic}".
		
		Your task is to formulate a highly effective Boolean query in MEDLINE format for PubMed.
		
		The query should balance \textbf{high recall} (capturing all relevant studies) with \textbf{reasonable precision} (avoiding irrelevant results):
		
		- Use both free-text terms and MeSH terms (e.g., chronic pain[tiab], Pain[mh]).
		
		- \textbf{Do not wrap terms or phrases in double quotes}, as this disables automatic term mapping (ATM).
		
		- Combine synonyms or related terms within a concept using OR.
		
		- Combine different concepts using AND.
		
		- Use wildcards (*) to capture word variants (e.g., vaccin* -> vaccine, vaccination):
		
		- Terms must have >= 4 characters before the * (e.g., colo*)
		
		- Wildcards work with field tags (e.g., breastfeed*[tiab]).
		
		- Field tags limit the search to specific fields and disable ATM.
		
		- Do not include date limits.
		
		- Tag term using term field (e.g., covid-19[ti] vaccine[ti] children[ti]) when needed.
		
		\textbf{Only use the following allowed field tags:}
		
		Title: [ti], Abstract: [ab], Title/Abstract: [tiab]
		
		MeSH: [mh], Major MeSH: [majr], Supplementary Concept: [nm]
		
		Text Words: [tw], All Fields: [all]
		
		Publication Type: [pt], Language: [la]

		Output and only output the formulated Boolean query inside <answer></answer> tags. Do not include any explanation or content outside or inside the <answer> tags.
	\end{PromptBox}
	
	\caption{No-reasoning prompt}
	\label{fig:no-reasoning-prompt}
\end{figure*}

%% file: tables/free-reasoning-prompt.tex
\begin{figure*}[ht!]
	\centering
	\small
	\begin{PromptBox}[System Message]
		You are an expert systematic review information specialist.
		
		You are tasked to formulate a systematic review Boolean query in response to a research topic.
		Your reasoning process should be enclosed within <think></think>, and the final Boolean query must be enclosed within <answer></answer> tags. Do not include anything outside of these tags.
	\end{PromptBox}
	
	\vspace{1em}
	
	\begin{PromptBox}[User Message]
		You are given a systematic review research topic, with the topic title "{topic}".
		
		Your task is to generate a highly effective Boolean query in MEDLINE format for PubMed.
		
		The query should balance \textbf{high recall} (capturing all relevant studies) with \textbf{reasonable precision} (avoiding irrelevant results):
		
		- Use both free-text terms and MeSH terms (e.g., chronic pain[tiab], Pain[mh]).
		
		- \textbf{Do not wrap terms or phrases in double quotes}, as this disables automatic term mapping (ATM).
		
		- Combine synonyms or related terms within a concept using OR.
		
		- Combine different concepts using AND.
		
		- Use wildcards (*) to capture word variants (e.g., vaccin* -> vaccine, vaccination):
		
		- Terms must have >= 4 characters before the * (e.g., colo*)
		
		- Wildcards work with field tags (e.g., breastfeed*[tiab]).
		
		- Field tags limit the search to specific fields and disable ATM.
		
		- Do not include date limits.
		
		- Tag terms using appropriate fields (e.g., covid-19[ti] vaccine[ti] children[ti]) when needed.
		
		\textbf{Only use the following allowed field tags:}
		
		Title: [ti], Abstract: [ab], Title/Abstract: [tiab]
		
		MeSH: [mh], Major MeSH: [majr], Supplementary Concept: [nm]
		
		Text Words: [tw], All Fields: [all]
		
		Publication Type: [pt], Language: [la]
		
		Output your full reasoning inside <think></think>.
		
		Output the final Boolean query inside <answer></answer>.
		
		Do not include any content outside these tags.
	\end{PromptBox}
	
	\caption{Free-text Reasoning prompt with <think> and <answer> outputs}
	\label{fig:reasoning-prompt}
\end{figure*}

%% file: tables/conceptual-reasoning-prompt.tex
\begin{figure*}[ht!]
	\centering
		\small
	
	\begin{PromptBox}[System Message]
		You are an expert systematic review information specialist.
		
		Formulate a systematic review Boolean query using step-by-step reasoning inside <think> </think>, and output the final query inside <answer> </answer>.
	\end{PromptBox}
	
	\vspace{1em}
	
	\begin{PromptBox}[User Message]
		You are given a systematic review topic titled: "{topic}".
		
		Construct a Boolean query using the \textbf{conceptual method}, based on domain logic and structured thinking.
		
		\textbf{Step 1}: Identify 2–3 key concepts from the topic (e.g., Population, Intervention, Outcome).
		
		\textbf{Step 2}: For each concept:
		- List related terms: synonyms, variants, relevant MeSH terms.
		- Prioritise specific, high-precision terms.
		
		\textbf{Step 3}: Create a Boolean block per concept:
		- Combine terms using OR
		- Use free-text terms and MeSH terms (e.g., chronic pain[tiab], Pain[mh])
		- \textbf{Do not wrap terms or phrases in double quotes}, as this disables automatic term mapping (ATM)
		- Tag terms individually when needed (e.g., covid-19[ti] vaccine[ti] children[ti])
		- Field tags limit search scope and disable ATM
		
		\textbf{Step 4}: Use wildcards (*) to capture word variants (e.g., vaccin* -> vaccine, vaccination):
		- Terms must have >= 4 characters before the * (e.g., colo*)
		- Wildcards work with field tags (e.g., breastfeed*[tiab]).
		
		\textbf{Step 5}: Combine all Boolean blocks using AND:
		((Concept1\_term1[tiab] OR Concept1\_term2[tiab] OR Concept1\_termX[mh]) AND (Concept2\_...))
		
		\textbf{Only use the following allowed field tags:}
		Title: [ti], Abstract: [ab], Title/Abstract: [tiab]
		MeSH: [mh], Major MeSH: [majr], Supplementary Concept: [nm]
		Text Words: [tw], All Fields: [all]
		Publication Type: [pt], Language: [la]
		
		Output your full reasoning inside <think>...</think>
		
		Output only the final Boolean query inside <answer>...</answer>
		
		Do not include any content outside these tags.
		
		Do not include date limits.
	\end{PromptBox}
	
	\caption{Conceptual-Reasoning prompt with <think> reasoning and <answer> output}
	\label{fig:prompt-conceptual}
\end{figure*}

%% file: tables/objective-reasoning-prompt.tex
\begin{figure*}[ht!]
	\centering
		\small
	\begin{PromptBox}[System Message]
		You are an expert systematic review information specialist.
		
		You are tasked to formulate a systematic review Boolean query step by step as a reasoning process within <think> </think>, and provide the Boolean query formulated <answer> </answer>.
	\end{PromptBox}
	
	\vspace{1em}
	
	\begin{PromptBox}[User Message]
		You are given a systematic review research topic, with the topic title "{topic}".
		
		You need to simulate a Boolean query construction process using the \textbf{objective method}, which is grounded in domain expertise and structured logic.
		
		\textbf{Step 1}: Simulate a concise title and abstract (2–3 sentences) of a \textit{relevant and focused} article clearly aligned with the topic. This is a hypothetical but plausible example.
		
		\textbf{Step 2}: Based on the simulated text, identify \textit{key informative terms or phrases} that best represent the article's core concepts. Prioritise specificity and informativeness. Avoid overly broad or ambiguous terms.
		
		\textbf{Step 3}: Categorise each term into one of the following:
		- (A) Health conditions or populations (e.g., diabetes, adolescents)
		- (B) Treatments, interventions, or exposures (e.g., insulin therapy, air pollution)
		- (C) Study designs or methodologies (e.g., randomized controlled trial, cohort study)
		- (N/A) Not applicable to any of the above categories
		
		\textbf{Step 4}: Using the categorised terms, build a Boolean query in MEDLINE format for PubMed:
		- Combine synonyms or related terms within each category using OR
		- Use both free-text terms and MeSH terms (e.g., chronic pain[tiab], Pain[mh])
		- \textbf{Do not wrap terms or phrases in double quotes}, as this disables automatic term mapping (ATM)
		- Tag each term individually when needed (e.g., covid-19[ti] vaccine[ti] children[ti])
		- Field tags limit the search to specific fields and disable ATM
		
		\textbf{Step 5}: Use wildcards (*) to capture word variants (e.g., vaccin* -> vaccine, vaccination):
		- Terms must have >= 4 characters before the * (e.g., colo*)
		- Wildcards work with field tags (e.g., breastfeed*[tiab]).
		
		\textbf{Step 6}: Combine all category blocks using AND:
		((itemA1[tiab] OR itemA2[tiab] OR itemA3[mh]) AND (itemB1[tiab] OR ...) AND (itemC1[tiab] OR ...))
		
		\textbf{Only use the following allowed field tags:}
		Title: [ti], Abstract: [ab], Title/Abstract: [tiab]
		MeSH: [mh], Major MeSH: [majr], Supplementary Concept: [nm]
		Text Words: [tw], All Fields: [all]
		Publication Type: [pt], Language: [la]
		
		Place your full reasoning (including simulated abstract, term list, classification, and query construction) inside <think></think>.
		
		Output the final Boolean query inside <answer></answer>.
		
		Do not include anything outside the <think> and <answer> tags.
		
		Do not include date restrictions.
	\end{PromptBox}
	
	\caption{Objective Reasoning prompt with simulated article and structured reasoning in <think>}
	\label{fig:prompt-objective}
\end{figure*}

%% file: tables/tunning-parameter.tex
\begin{table}[h]
	\centering
	\small
		\caption{Training hyperparameters used for GRPO-based Boolean query generation.}
	\begin{tabular}{ll}
		\toprule
		\textbf{Parameter} & \textbf{Value} \\
		\midrule
		Adapter type            & LoRA \\
		LoRA rank ($r$)         & 16 \\
		LoRA alpha              & 32 \\
		LoRA dropout            & 0.05 \\
		Quantization            & bf16 \\
		Attn implementation     & flash-attention-2 \\
		Effective batch size    & 16 \\
		Learning rate           & 1e-5 \\
		Generation temperature  & 0.6 \\
		\# Generations / prompt & 4 \\
		Prompt length (non-reasoning)   & 768 / 1024 tokens \\
		Prompt length (reasoning-based) & 1024 / 3072 tokens \\
		Reward functions        & Format, Validity, Retrieval \\
		Inference engine        & vLLM (colocate mode) \\
		Optimizer backend       & DeepSpeed \\
		\# Epochs               & 1 \\
		\bottomrule
	\end{tabular}

	\label{appendix:tab-training-details}
\end{table}

%% file: tables/impact-retrieval-reward.tex
\begin{table*}[t]

	\centering
	\caption{Impact of removing individual components from the AutoBool retrieval reward design.}
	\label{tab:eval-results-reward-components}
		\resizebox{\linewidth}{!}{
	\begin{tabular}{llcccc}
		\toprule
		Prompt & Model & Recall & F3 & Recall > 80\% & Recall > 90\% \\
		\midrule
		\multirow{4}{*}{N.R} & AutoBool & 0.7036 & \textbf{0.1195} & 47.10 & 32.30 \\
		& W/O Scaling & \textbf{0.7627} (+8.4\%) & 0.0429 (-64.1\%) & \textbf{56.60} (+20.2\%) & \textbf{40.90} (+26.6\%) \\
		& W/O Dependency & 0.4560 (-35.2\%) & 0.1191 (-0.3\%) & 18.20 (-61.4\%) & 10.50 (-67.5\%) \\
		& W/O Precision & 0.6419 (-8.8\%) & 0.0850 (-28.8\%) & 40.00 (-15.1\%) & 28.20 (-12.7\%) \\
		\midrule
		\multirow{4}{*}{R} & AutoBool & 0.5453 & \textbf{0.1346} & 27.90 & 16.50 \\
		& W/O Scaling & 0.5129 (-5.9\%) & 0.1244 (-7.6\%) & 22.20 (-20.4\%) & 13.00 (-21.2\%) \\
		& W/O Dependency & \textbf{0.6367} (+16.8\%) & 0.1028 (-23.6\%) & \textbf{37.20} (+33.3\%) & \textbf{24.60} (+49.1\%) \\
		& W/O Precision & 0.4880 (-10.5\%) & 0.1185 (-12.0\%) & 21.30 (-23.7\%) & 13.00 (-21.2\%) \\
		\midrule
		\multirow{4}{*}{R-con} & AutoBool & 0.5262 & \textbf{0.1066} & 26.30 & 16.80 \\
		& W/O Scaling & 0.5403 (+2.7\%) & 0.0799 (-25.1\%) & 26.80 (+1.9\%) & \textbf{16.90} (+0.6\%) \\
		& W/O Dependency & \textbf{0.5484} (+4.2\%) & 0.0847 (-20.6\%) & 26.30 (+0.0\%) & 16.60 (-1.2\%) \\
		& W/O Precision & 0.5426 (+3.1\%) & 0.0939 (-11.9\%) & \textbf{27.70} (+5.3\%) & 15.60 (-7.1\%) \\
		\midrule
		\multirow{4}{*}{R-obj} & AutoBool & 0.6540 & 0.1094 & \textbf{39.70} & 26.70 \\
		& W/O Scaling & 0.6322 (-3.3\%) & 0.0848 (-22.5\%) & 36.40 (-8.3\%) & 24.10 (-9.7\%) \\
		& W/O Dependency & \textbf{0.6669} (+2.0\%) & \textbf{0.1191} (+8.9\%) & 39.30 (-1.0\%) & \textbf{27.20} (+1.9\%) \\
		& W/O Precision & 0.6385 (-2.4\%) & 0.0805 (-26.5\%) & 38.10 (-4.0\%) & 24.40 (-8.6\%) \\
		\bottomrule
	\end{tabular}
}
\end{table*}

%% file: tables/impact-f3.tex
\begin{table*}[t]
	
	\centering
	\caption{Comparison between AutoBool's original retrieval reward versus direct F3 optimization.}
	\label{tab:eval-results-reward-f3}
	\resizebox{\linewidth}{!}{
	\begin{tabular}{llcccc}
		\toprule
		Prompt & Model & Recall & F3 & Recall > 80\% & Recall > 90\% \\
		\midrule
		\multirow{2}{*}{N.R} & AutoBool & \textbf{0.7036} & 0.1195 & \textbf{47.10} & \textbf{32.30} \\
		& F3 Optimization & 0.3656 (-48.0\%) & \textbf{0.1463} (+22.5\%) & 12.00 (-74.5\%) & 6.80 (-78.9\%) \\
		\midrule
		\multirow{2}{*}{R} & AutoBool & \textbf{0.5453} & 0.1346 & \textbf{27.90} & \textbf{16.50} \\
		& F3 Optimization & 0.4051 (-25.7\%) & \textbf{0.1547} (+15.0\%) & 13.50 (-51.6\%) & 8.20 (-50.3\%) \\
		\midrule
		\multirow{2}{*}{R-con} & AutoBool & \textbf{0.5262} & 0.1066 & \textbf{26.30} & \textbf{16.80} \\
		& F3 Optimization & 0.4436 (-15.7\%) & \textbf{0.1186} (+11.2\%) & 17.90 (-31.9\%) & 11.60 (-31.0\%) \\
		\midrule
		\multirow{2}{*}{R-obj} & AutoBool & \textbf{0.6540} & 0.1094 & \textbf{39.70} & \textbf{26.70} \\
		& F3 Optimization & 0.3977 (-39.2\%) & \textbf{0.1891} (+72.8\%) & 12.50 (-68.5\%) & 6.70 (-74.9\%) \\
		\bottomrule
	\end{tabular}
}
\end{table*}

%% file: tables/impact-m.tex
\begin{table}[t]
	\small
	\centering
	\caption{Impact of scaling factor $M$ on trained AutoBool model effectiveness.}
	\label{tab:eval-results-m-values}

	\begin{tabular}{llcccc}
		\toprule
		 & M& Recall & F3 & \parbox{2.5em}{Recall\\>80\%}  & \parbox{2.5em}{Recall\\>90\%}  \\
		\midrule
		\multirow{3}{*}{\rotatebox{90}{N.R}} & 5 & 0.4368 & 0.1119 & 18.20 & 10.80 \\
		& 10 & \textbf{0.7036} & \textbf{0.1195} & \textbf{47.10} & \textbf{32.30} \\
		& 20 & 0.5524 & 0.1182 & 28.50 & 18.60 \\
		\midrule
		\multirow{3}{*}{\rotatebox{90}{R}} & 5 & 0.5167 & \textbf{0.1349} & 23.70 & 15.10 \\
		& 10 & 0.5453 & 0.1346 & 27.90 & 16.50 \\
		& 20 & \textbf{0.5817} & 0.1155 & \textbf{31.30} & \textbf{20.40} \\
		\midrule
		\multirow{3}{*}{\rotatebox{90}{R-con}} & 5 & 0.5301 & 0.0981 & 24.00 & 14.70 \\
		& 10 & 0.5262 & \textbf{0.1066} & \textbf{26.30} & \textbf{16.80} \\
		& 20 & \textbf{0.5302} & 0.0893 & 24.90 & 14.80 \\
		\midrule
		\multirow{3}{*}{\rotatebox{90}{R-obj}} & 5 & 0.6322 & 0.1059 & 35.80 & 23.80 \\
		& 10 & 0.6540 & 0.1094 & 39.70 & 26.70 \\
		& 20 & \textbf{0.6722} & \textbf{0.1204} & \textbf{42.70} & \textbf{27.90} \\
		\bottomrule
	\end{tabular}

\end{table}

%% file: tables/impact-s.tex
\begin{table}[t]
	\small
	\centering
	\caption{Impact of smoothing constant $s$ on trained AutoBool model effectiveness.}
	\label{tab:eval-results-scale-values}

	\begin{tabular}{llcccc}
		\toprule
		 & s & Recall & F3 & \parbox{2.5em}{Recall\\>80\%}  & \parbox{2.5em}{Recall\\>90\%} \\
		\midrule
		\multirow{3}{*}{\rotatebox{90}{N.R}} & 10 & 0.3949 & 0.1065 & 15.50 & 9.20 \\
		& 100 & \textbf{0.7036} & 0.1195 & \textbf{47.10} & \textbf{32.30} \\
		& 1000 & 0.4845 & \textbf{0.1242} & 22.20 & 13.30 \\
		\midrule
		\multirow{3}{*}{\rotatebox{90}{R}} & 10 & \textbf{0.5738} & 0.1172 & 28.90 & 18.10 \\
		& 100 & 0.5453 & \textbf{0.1346} & 27.90 & 16.50 \\
		& 1000 & 0.5592 & 0.1274 & \textbf{29.10} & \textbf{19.00} \\
		\midrule
		\multirow{3}{*}{\rotatebox{90}{R-con}} & 10 & 0.5461 & 0.0909 & 27.10 & 16.40 \\
		& 100 & 0.5262 & \textbf{0.1066} & 26.30 & 16.80 \\
		& 1000 & \textbf{0.5761} & 0.0887 & \textbf{28.20} & \textbf{18.70} \\
		\midrule
		\multirow{3}{*}{\rotatebox{90}{R-obj}} & 10 & \textbf{0.6793} & 0.0852 & \textbf{43.40} & \textbf{29.00} \\
		& 100 & 0.6540 & \textbf{0.1094} & 39.70 & 26.70 \\
		& 1000 & 0.6440 & 0.1079 & 39.10 & 25.10 \\ \bottomrule
		
	\end{tabular}

\end{table}

%% file: tables/impact-retrieval-panelty.tex
\begin{table}[t]
	\small
	\centering
	\caption{Impact of empty result penalty values on trained AutoBool model effectiveness.}
	\label{tab:eval-results-min-reward}

		\begin{tabular}{llcccc}
			\toprule
			& Penalty & Recall & F3 & \parbox{2.5em}{Recall\\>80\%}  & \parbox{2.5em}{Recall\\>90\%} \\
			\midrule
			\multirow{3}{*}{\rotatebox{90}{N.R}} & -10 & 0.5268 & 0.1125 & 26.80 & 17.70 \\
			& -20 & \textbf{0.7036} & 0.1195 & \textbf{47.10} & \textbf{32.30} \\
			& -30 & 0.4618 & \textbf{0.1220} & 22.10 & 13.90 \\
			\midrule
			\multirow{3}{*}{\rotatebox{90}{R}} & -10 & \textbf{0.5576} & \textbf{0.1360} & 27.40 & \textbf{19.10} \\
			& -20 & 0.5453 & 0.1346 & \textbf{27.90} & 16.50 \\
			& -30 & 0.5200 & 0.1278 & 23.20 & 15.10 \\
			\midrule
			\multirow{3}{*}{\rotatebox{90}{R-con}} & -10 & 0.5380 & 0.0928 & 25.30 & 16.60 \\
			& -20 & 0.5262 & \textbf{0.1066} & \textbf{26.30} & \textbf{16.80} \\
			& -30 & \textbf{0.5516} & 0.0817 & 25.60 & \textbf{16.80} \\
			\midrule
			\multirow{3}{*}{\rotatebox{90}{R-obj}} & -10 & 0.6040 & \textbf{0.1210} & 32.30 & 20.70 \\
			& -20 & \textbf{0.6540} & 0.1094 & \textbf{39.70} & \textbf{26.70} \\
			& -30 & 0.6521 & 0.1056 & 39.20 & 25.10 \\
			\bottomrule
		\end{tabular}
	
\end{table}

\begin{table}[t]
	\small
	\centering
	\caption{Impact of zero relevant results penalty values on trained AutoBool model effectiveness.}
	\label{tab:eval-results-off-topic-penalty}

	\begin{tabular}{llcccc}\toprule
		& Penalty & Recall & F3 & \parbox{2.5em}{Recall\\>80\%}  & \parbox{2.5em}{Recall\\>90\%} \\
		\midrule
		\multirow{3}{*}{\rotatebox{90}{N.R}} & 0 & 0.4183 & 0.1144 & 16.60 & 9.60 \\
		& -5 & \textbf{0.7036} & \textbf{0.1195} & \textbf{47.10} & \textbf{32.30} \\
		& -10 & 0.5072 & 0.1034 & 24.00 & 15.30 \\
		\midrule
		\multirow{3}{*}{\rotatebox{90}{R}} & 0 & 0.4809 & 0.1276 & 20.70 & 11.90 \\
		& -5 & \textbf{0.5453} & \textbf{0.1346} & \textbf{27.90} & \textbf{16.50} \\
		& -10 & 0.5416 & 0.1318 & 25.60 & 16.30 \\
		\midrule
		\multirow{3}{*}{\rotatebox{90}{R-con}} & 0 & \textbf{0.5469} & 0.0921 & \textbf{26.70} & 16.40 \\
		& -5 & 0.5262 & \textbf{0.1066} & 26.30 & \textbf{16.80} \\
		& -10 & 0.5456 & 0.0887 & 26.10 & 15.60 \\
		\midrule
		\multirow{3}{*}{\rotatebox{90}{R-obj}} & 0 & 0.6288 & 0.1040 & 37.10 & 23.50 \\
		& -5 & \textbf{0.6540} & \textbf{0.1094} & \textbf{39.70} & \textbf{26.70} \\
		& -10 & 0.6384 & 0.1066 & 36.10 & 23.70 \\
		
		\bottomrule
	\end{tabular}

\end{table}

%% file: tables/example.tex
\newtcolorbox{CombinedBox}[1][]{%
	breakable,
	colback=gray!5,
	colframe=black!40,
	title=#1,
	fonttitle=\bfseries,
	boxrule=0.4pt,
	arc=2pt,
	top=1pt,
	bottom=1pt,
	left=2pt,
	right=2pt,
	width=\textwidth,
}

\begin{figure*}[t]
	\centering
	
	\begin{CombinedBox}[No-Reasoning]
		\textbf{\small Systematic Review Topic}
		
		\vspace{0.3em}
		{\footnotesize
			Topic ID: 39707254 \\ The asymptomatic tuberculosis proportion among the elderly population
			\\
			\textbf{Max Publication Date:} 2024/12/20 \quad \textbf{Number of Ground Truth Documents:} 28
		}
		
		\vspace{0.5em}
		\hrule
		\vspace{0.5em}
		
	\textbf{\small Generated Response}
		
		\vspace{0.3em}
		\begin{lstlisting}[basicstyle=\ttfamily\scriptsize, breaklines=true, breakatwhitespace=true, showstringspaces=false, columns=fullflexible, keepspaces=true, breakindent=0pt, frame=none, xleftmargin=0pt, xrightmargin=0pt]
<answer>
((tuberculosis[tx] OR TB[tx] OR Mycobacterium tuberculosis[tx] OR TB[mh]) AND (elderly[tx] OR elderly[mh] OR geriatric[tx] OR geriatrics[tx] OR aged[tx] OR older adult[tx])) AND ((asymptomatic[tx] OR asymptomatic[mh] OR undiagnosed[tx] OR latent[tx] OR latent TB[tx]) OR (prevalence[tx] OR incidence[tx] OR prevalence[mh] OR incidence[mh] OR prevalence of tuberculosis[tx] OR TB testing[tx] OR TB screening[tx]))
</answer>
		\end{lstlisting}
		
		\vspace{0.5em}
		\hrule
		\vspace{0.5em}
		
		\textbf{\small Performance Metrics}
		
		\vspace{0.3em}
		{\small
			\begin{tabular}{@{}lcccc@{}}
				\toprule
				\textbf{Metric} & Precision & F1 & F3 & Recall \\
				\midrule
				\textbf{Value} & 0.0001 & 0.0002 & 0.0012 & 1.0000 \\
				\bottomrule
			\end{tabular}
		}
	\end{CombinedBox}
	
	\caption{Results with no reasoning for topic ID 39707254.}
	\label{fig:topic_39707254_no_reasoning}
\end{figure*}

% Figure 2: Reasoning

%% file: tables/example-b.tex
\begin{figure*}[p]
	\centering
	
	\begin{CombinedBox}[Free-text Reasoning]
		\textbf{\small Systematic Review Topic}
		
		\vspace{0.3em}
		{\footnotesize
			Topic ID: 39707254 \\ The asymptomatic tuberculosis proportion among the elderly population
			\\
			\textbf{Max Publication Date:} 2024/12/20 \quad \textbf{Number of Ground Truth Documents:} 28
		}
		
		\vspace{0.5em}
		\hrule
		\vspace{0.5em}
		
		\textbf{\small Generated Response}
		
		\vspace{0.3em}
		\begin{lstlisting}[basicstyle=\ttfamily\scriptsize, breaklines=true, breakatwhitespace=true, showstringspaces=false, columns=fullflexible, keepspaces=true, breakindent=0pt, frame=none, xleftmargin=0pt, xrightmargin=0pt]
<think>
Okay, let's tackle this. The user wants a Boolean query for a systematic review on the asymptomatic TB proportion among the elderly. First, I need to identify key terms related to tuberculosis and the elderly. 

For TB, MeSH terms like "Tuberculosis" and "Tuberculosis, Pulmonary" are important. Also, "Tuberculosis, Type 2" might be relevant. Then, "asymptomatic" is crucial. I should check if "Asymptomatic" is a MeSH term. Maybe "Tuberculosis, Asymptomatic" is a possible MeSH term, but I'm not sure. Alternatively, using "asymptomatic" in the title or abstract.

Elderly population: "Elderly" is a MeSH term. Also, terms like "Aged" or "Geriatrics" might be relevant. Need to include both MeSH and free-text terms. 

Combining these, the query should have terms related to TB, asymptomatic, and elderly. Let me structure it. Use MeSH terms for TB and elderly. Then include free-text terms like "asymptomatic" and "elderly" in the title or abstract. Also, consider synonyms like "elderly" could be "Aged", "Older Adults", etc. 

Wildcards might be useful. For example, "tubercul*" could capture tuberculosis, tuberculous, etc. But need to ensure that the terms are properly formatted. Also, check if "proportion" or "prevalence" are relevant. Since the review is about proportion, terms like "proportion", "prevalence", "incidence" might be useful. However, the user specified to focus on the topic, so maybe stick to the main terms.

Putting it all together: 

MeSH terms for TB: Tuberculosis[MeSH], Tuberculosis, Pulmonary[MeSH], Tuberculosis, Type 2[MeSH]. 

MeSH term for elderly: Elderly[MeSH], Aged[MeSH].

Free-text terms: asymptomatic[tiab], elderly[tiab], proportion[tiab], prevalence[tiab], TB[tiab], tuberculosis[tiab].

Combine these with ANDs and ORs. For example, (Tuberculosis[MeSH] OR Tuberculosis[tiab]) AND (asymptomatic[tiab] OR asymptomatic[MeSH]) AND (Elderly[MeSH] OR Aged[MeSH] OR elderly[tiab]).

Wait, but the user said to use both MeSH and free-text. Also, need to check if "asymptomatic" is a MeSH term. Maybe "Asymptomatic" is a MeSH term, but I think it's more common to use free text. Alternatively, check if "Tuberculosis, Asymptomatic" is a MeSH term. If not, then use free text.

Also, need to include terms related to the study's focus, like "proportion" or "prevalence" in the abstract. Maybe include those in the title/abstract field.

So the final query would be something like:

(Tuberculosis[MeSH] OR Tuberculosis[tiab]) AND (asymptomatic[tiab] OR Asymptomatic[MeSH]) AND (Elderly[MeSH] OR Aged[MeSH] OR elderly[tiab]) AND (proportion[tiab] OR prevalence[tiab] OR incidence[tiab])

But need to make sure that all these terms are properly combined. Also, check for possible synonyms. For example, "elderly" could be "Older Adults", but maybe that's too specific. Also, "Aged" is a MeSH term, so that's good.

Another consideration: the query should be in MEDLINE format, so using the correct field tags. The user specified that only certain fields are allowed, like title, abstract, MeSH, etc. So the query should use these fields.

Putting it all together, the Boolean query would combine MeSH terms and free-text terms, using ORs where appropriate. Let me structure it step by step.
</think>

<answer>  
((Tuberculosis[MeSH] OR Tuberculosis[tiab] OR Tuberculos[tiab]) AND (Asymptomatic[tiab] OR Asymptomatic[MeSH] OR asymptomatic[tiab])) AND ((Elderly[MeSH] OR Aged[MeSH] OR Elderly[tiab] OR Aged[tiab] OR Older[tiab] OR Geriatric[tiab]))  
</answer>
		\end{lstlisting}
		
		\vspace{0.5em}
		\hrule
		\vspace{0.5em}
		
		\textbf{\small Performance Metrics}
		
		\vspace{0.3em}
		{\small
			\begin{tabular}{@{}lcccc@{}}
				\toprule
				\textbf{Metric} & Precision & F1 & F3 & Recall \\
				\midrule
				\textbf{Value} & 0.0038 & 0.0076 & 0.0347 & 0.3333 \\
				\bottomrule
			\end{tabular}
		}
	\end{CombinedBox}
	
	\caption{Results with free-text reasoning for topic ID 39707254.}
	\label{fig:topic_39707254_reasoning}
\end{figure*}

% Figure 3: Conceptual Reasoning
\begin{figure*}[p]
	\centering
	
	\begin{CombinedBox}[Conceptual Reasoning]
		\textbf{\small Systematic Review Topic}
		
		\vspace{0.3em}
		{\footnotesize
			Topic ID: 39707254 \\ The asymptomatic tuberculosis proportion among the elderly population
			\\
			\textbf{Publication Date:} 2024/12/20 \quad \textbf{Number of Ground Truth Documents:} 28
		}
		
		\vspace{0.5em}
		\hrule
		\vspace{0.5em}
		
		\textbf{\small Generated Response}
		
		\vspace{0.3em}
		\begin{lstlisting}[basicstyle=\ttfamily\scriptsize, breaklines=true, breakatwhitespace=true, showstringspaces=false, columns=fullflexible, keepspaces=true, breakindent=0pt, frame=none, xleftmargin=0pt, xrightmargin=0pt]
<think>
Okay, let's tackle this. The user wants a Boolean query for a systematic review on the asymptomatic tuberculosis proportion among the elderly. First, I need to identify the key concepts. The main topic is "asymptomatic tuberculosis" and the population is "elderly". So the two main concepts are "asymptomatic tuberculosis" and "elderly population".

For the first concept, "asymptomatic tuberculosis", I should think of synonyms and related terms. "Asymptomatic" could be "no symptoms" or "symptom-free". "Tuberculosis" is straightforward, but maybe "TB" is a shorter term. Also, MeSH terms might be relevant. The MeSH term for tuberculosis is "Tuberculosis" and "Tuberculosis, Type 2" might be specific. "Asymptomatic" in MeSH is "Asymptomatic" or "Symptom-Free".

For the second concept, "elderly population", terms like "elderly", "senior", "old", "aged", "elderly people". MeSH terms for age groups include "Elderly" and "Aged". Also, "Geriatric" might be relevant. The MeSH term for age is "Age Factors" or "Age Groups".

Now, structuring each concept into Boolean blocks. For "asymptomatic tuberculosis", I'll use terms like "asymptomatic", "symptom-free", "TB", "Tuberculosis", and MeSH terms. For the elderly, terms like "elderly", "aged", "senior", "old", and MeSH terms "Elderly", "Aged", "Age Factors".

I need to combine these with OR for each term and then AND the two concepts. Also, using field tags: title, abstract, MeSH. For example, "asymptomatic[tiab] OR symptom-free[tiab] OR tuberculosis[tiab] OR TB[tiab] OR Tuberculosis[tiab] OR asymptomatic[mh] OR symptom-free[mh]". Then for the elderly: "elderly[tiab] OR aged[tiab] OR senior[tiab] OR old[tiab] OR elderly[mh] OR aged[mh] OR age factors[mh]".

Combine both concepts with AND. Also, check if MeSH terms should be included in the right fields. The user mentioned using MeSH terms with [mh], so I'll include those. Also, make sure to use the allowed field tags: [tiab], [mh], [majr], etc. But the user said to use [tiab], [mh], [tw], [all], [pt], [la]. Since the query is for titles and abstracts, maybe [tiab] is sufficient, but including MeSH terms in [mh] as well.

Wait, the user said to use MeSH terms in [mh], so for each concept, include both free-text terms and MeSH terms. So for "asymptomatic tuberculosis", terms like "asymptomatic[tiab] OR asymptomatic[mh] OR symptom-free[tiab] OR symptom-free[mh] OR tuberculosis[tiab] OR TB[tiab] OR Tuberculosis[mh]". Similarly for the elderly.

Then, combine the two concepts with AND. So the full query would be:

(asymptomatic[tiab] OR asymptomatic[mh] OR symptom-free[tiab] OR symptom-free[mh] OR tuberculosis[tiab] OR TB[tiab] OR Tuberculosis[mh]) AND (elderly[tiab] OR elderly[mh] OR aged[tiab] OR aged[mh] OR senior[tiab] OR old[tiab] OR age factors[mh] OR age groups[mh])

I need to make sure that all terms are in the right fields. Also, check if "asymptomatic" is a MeSH term. Looking up, "Asymptomatic" is a MeSH term, so that's correct. "Tuberculosis" is the MeSH term. For the elderly, "Elderly" is a MeSH term. Also, "Aged" is another MeSH term. "Age Factors" is a MeSH term as well.

So the final query should include both the free-text terms and MeSH terms in their respective fields. Also, using [tiab] for title and abstract, and [mh] for MeSH terms. The user specified to use OR for each term within each concept, then AND between concepts. That should cover the necessary terms.
</think>

<answer>
((asymptomatic[tiab] OR asymptomatic[mh] OR symptom-free[tiab] OR symptom-free[mh] OR tuberculosis[tiab] OR TB[tiab] OR Tuberculosis[mh]) AND (elderly[tiab] OR elderly[mh] OR aged[tiab] OR aged[mh] OR senior[tiab] OR old[tiab] OR age factors[mh] OR age groups[mh]))
</answer>
		\end{lstlisting}
		
		\vspace{0.5em}
		\hrule
		\vspace{0.5em}
		
		\textbf{\small Performance Metrics}
		
		\vspace{0.3em}
		{\small
			\begin{tabular}{@{}lcccc@{}}
				\toprule
				\textbf{Metric} & Precision & F1 & F3 & Recall \\
				\midrule
				\textbf{Value} & 0.0000 & 0.0000 & 0.0002 & 1.0000 \\
				\bottomrule
			\end{tabular}
		}
	\end{CombinedBox}
	
	\caption{Results with conceptual reasoning for topic ID 39707254.}
	\label{fig:topic_39707254_conceptual_reasoning}
\end{figure*}

% Figure 4: Objective Reasoning
\begin{figure*}[p]
	\centering
	
	\begin{CombinedBox}[Objective Reasoning]
		\textbf{\small Systematic Review Topic}
		
		\vspace{0.3em}
		{\footnotesize
			Topic ID: 39707254 \\ The asymptomatic tuberculosis proportion among the elderly population
			\\
			\textbf{Max Publication Date:} 2024/12/20 \quad \textbf{Number of Ground Truth Documents:} 28
		}
		
		\vspace{0.5em}
		\hrule
		\vspace{0.5em}
		
		\textbf{\small Generated Response}
		
		\vspace{0.3em}
		\begin{lstlisting}[basicstyle=\ttfamily\scriptsize, breaklines=true, breakatwhitespace=true, showstringspaces=false, columns=fullflexible, keepspaces=true, breakindent=0pt, frame=none, xleftmargin=0pt, xrightmargin=0pt]
<think>
Okay, so I need to create a Boolean query for a systematic review on the asymptomatic tuberculosis proportion among the elderly. Let me start by understanding the topic. The main focus is on identifying studies that look at how many elderly people have TB without symptoms. 

First, I should simulate a relevant article's title and abstract. Maybe something like "Prevalence of Asymptomatic Tuberculosis in Elderly Patients: A Cross-Sectional Study in Urban Settings." The abstract would mention methods like cohort studies, population-based data, and statistical analysis to determine the proportion.

Next, identifying key terms. The main concepts here are "asymptomatic tuberculosis," "elderly population," and "study design." Terms like "tuberculosis," "asymptomatic," "elderly," "cohort study," "prevalence," "incidence," "cross-sectional study," and "elderly" are important. Also, MeSH terms like "Tuberculosis, Pulmonary" and "Elderly" would be relevant.

Categorizing these terms: 
- Health conditions or populations: tuberculosis, elderly
- Study designs: cohort study, cross-sectional study
- Treatments or interventions: not applicable here since it's about prevalence, not treatment

Now, building the Boolean query. I need to combine terms within each category using OR and then AND them together. For the title and abstract fields, I'll use [tiab]. MeSH terms can be added with [mh] or [majr]. For example, "tuberculosis[tiab] OR tuberculosis[mh]" to cover both terms. 

Wildcards might be useful for terms like "tubercul*" but need to ensure they have at least four characters before the asterisk. However, "tuberculosis" is specific enough, so maybe not necessary. 

Putting it all together: 
- Title/Abstract fields for terms related to TB and elderly.
- MeSH terms for tuberculosis and elderly.
- Study designs like cohort study, cross-sectional study, or case-control.

I should structure it as ((tuberculosis[tiab] OR tuberculosis[mh]) AND (elderly[tiab] OR elderly[mh]) AND (cohort study[tiab] OR cross-sectional study[tiab] OR prevalence[tiab] OR incidence[tiab] OR cross-sectional[tiab] OR cohort[tiab] OR study[tiab])).

Wait, maybe include "asymptomatic" as a key term. So adding "asymptomatic[tiab] OR asymptomatic[mh]" to narrow down the search. That would help target studies specifically looking at asymptomatic cases.

Also, check if "elderly" is a MeSH term. Yes, "Elderly" is a MeSH term. So using that in [mh] would be better. 

Putting all these elements together, the final query would combine all these components using AND to ensure all categories are met. Let me make sure each part is correctly formatted with the allowed field tags.
</think>

<answer>  
((tuberculosis[tiab] OR tuberculosis[mh] OR TB[tiab] OR tuberculosis[all]) AND (elderly[tiab] OR elderly[mh] OR older adults[tiab] OR elderly[all]) AND (cohort study[tiab] OR cross-sectional study[tiab] OR prevalence[tiab] OR incidence[tiab] OR asymptomatic[tiab] OR asymptomatic tuberculosis[tiab] OR asymptomatic[all]))  
</answer>
		\end{lstlisting}
		
		\vspace{0.5em}
		\hrule
		\vspace{0.5em}
		
		\textbf{\small Performance Metrics}
		
		\vspace{0.3em}
		{\small
			\begin{tabular}{@{}lcccc@{}}
				\toprule
				\textbf{Metric} & Precision & F1 & F3 & Recall \\
				\midrule
				\textbf{Value} & 0.0002 & 0.0004 & 0.0019 & 0.6667 \\
				\bottomrule
			\end{tabular}
		}
	\end{CombinedBox}
	
	\caption{Results with objective reasoning for topic ID 39707254.}
	\label{fig:topic_39707254_objective_reasoning}
\end{figure*}